\documentclass{SciPost}
\usepackage{amsmath,amscd,amsthm,latexsym,euscript}
\usepackage{tikz}
\usetikzlibrary{cd}
\usepackage{subcaption}
\DeclareCaptionLabelFormat{itletter}{(\textit{#2})}
\hypersetup{
    colorlinks,
    linkcolor={red!50!black},
    citecolor={blue!50!black},
    urlcolor={blue!80!black}
}

\usepackage[bitstream-charter]{mathdesign}
\DeclareSymbolFont{usualmathcal}{OMS}{cmsy}{m}{n}
\DeclareSymbolFontAlphabet{\mathcal}{usualmathcal}

\newtheorem{theor}{Theorem}

\theoremstyle{definition}

\newtheorem{prop}[theor]{Proposition}
\newtheorem{lem}[theor]{Lemma}
\newtheorem{define}{Definition}

\newtheorem{problem}{Problem}
\newtheorem{open}[problem]{Open problem}

\theoremstyle{remark}
\newtheorem{rem}{Remark}

\theoremstyle{definition}
\theoremstyle{definition}

\newcommand{\BBR}{\mathbb{R}}\newcommand{\BBC}{\mathbb{C}}
 
\newcommand{\BBS}{\mathbb{S}}
\newcommand{\BBZ}{\mathbb{Z}}

\newcommand{\cB}{\mathcal{B}}

\newcommand{\boldb}{{\boldsymbol{b}}}
\newcommand{\Bone}{{\boldsymbol{1}}}
\newcommand{\ba}{{\boldsymbol{a}}}

\newcommand{\bc}{{\boldsymbol{c}}}

\newcommand{\be}{{\boldsymbol{e}}}

\newcommand{\bn}{{\boldsymbol{n}}}
\newcommand{\bm}{{\boldsymbol{m}}}

\newcommand{\bu}{{\boldsymbol{u}}}
\newcommand{\bv}{{\boldsymbol{v}}}
\newcommand{\bw}{{\boldsymbol{w}}}
\newcommand{\bx}{{\boldsymbol{x}}}

\newcommand{\btheta}{{\boldsymbol{\theta}}}
\newcommand{\bepsilon}{{\boldsymbol{\epsilon}}}

\newcommand{\ceq}{\mathrel{{:}{=}}}

\DeclareMathOperator{\Area}{Area}
\DeclareMathOperator{\Vol}{Vol}
\DeclareMathOperator{\Span}{span}

\DeclareMathOperator{\id}{id}

\DeclareMathOperator{\tr}{tr}

\newcommand{\su}{\mathfrak{su}}

\DeclareMathOperator{\Real}{Re}
\DeclareMathOperator{\Imag}{Im}

\newcommand{\wedges}{\wedge\cdots\wedge}

\newcommand{\vol}[1]{\textbf{{#1}}}

\newcommand{\PP}[2]{({\mx+(#1)*\eax+(#2)*\ebx},{\my+(#1)*\eay+(#2)*\eby},{\mz+(#1)*\eaz+(#2)*\ebz})}
\newcommand{\QQT}[3]{({\mxT+(#1)*\eaxT+(#2)*\ebxT+(#3)*\nxT},{\myT+(#1)*\eayT+(#2)*\ebyT+(#3)*\nyT},{\mzT+(#1)*\eazT+(#2)*\ebzT+(#3)*\nzT})}
\newcommand{\PPT}[2]{\QQT{#1}{#2}{0}}

\fancypagestyle{SPstyle}{
\fancyhf{}
\lhead{\colorbox{scipostblue}{\bf \color{white} ~SciPost Physics Proceedings }}
\rhead{{\bf \color{scipostdeepblue} ~Submission }}

\fancyfoot[C]{\textbf{\thepage}}
}
\begin{document}

\pagestyle{SPstyle}

\begin{center}{\Large \textbf{\color{scipostdeepblue}{
The Wronskian determinant $W_{d=3}^{k=1}$ over $\mathbb{R}^3$ calculates\\ the baryon current charge density $\mathcal{B}^0$ of the Skyrme model
}}}\end{center}

\begin{center}\textbf{
Markuss G. \c{K}\={e}ni\c{n}\v{s}\textsuperscript{1,2$\star$} and
Arthemy V{}. Kiselev\textsuperscript{2$\dagger$}
}\end{center}

\begin{center}
{\bf 1} ETH Z\"urich -- Department of Mathematics, R\"{a}mistrasse 101, CH-8092 Z\"{u}rich, Switzerland.
\\
{\bf 2} Bernoulli Institute for Mathematics, Computer Science \&\ 
Artificial Intelligence,\\ 
University of Groningen, P.O.\,Box\:407, 9700\,AK Groningen, 
The Netherlands.
\\[\baselineskip]
$\star$ \href{mailto:mkenins@student.ethz.ch}{\small mkenins@student.ethz.ch}\,,\quad
$\dagger$ \href{mailto:a.v.kiselev@rug.nl}{\small a.v.kiselev@rug.nl}
\end{center}

\definecolor{palegray}{gray}{0.95}
\begin{center}
\colorbox{palegray}{
  \begin{tabular}{ll}
 \hspace{-11pt} \begin{minipage}{0.18\textwidth}
    \includegraphics[width=28mm]{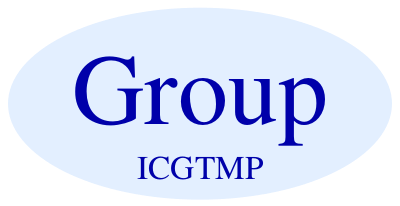}
  \end{minipage}
  &
  \begin{minipage}{0.75\textwidth}
    \begin{center}
    {\it 36th International Colloquium on Group Theoretical Methods in Physics}\\
    {\it Valladolid, 13-17 July 2026} \\
    \doi{10.21468/SciPostPhysProc.?}\\
    \end{center}
  \end{minipage}
\end{tabular}
}
\end{center}

\section*{\color{scipostdeepblue}{Abstract}}
\textbf{\boldmath{%
The Skyrme model is described by using Jacobians of topologically non-trivial maps $U\colon \mathbb{R}^3\to SU(2)$; we re-describe the model in terms of generalised Wronskians over $\mathbb{R}^3$. We exploit the isomorphism $SU(2)\cong \mathbb{S}^3$, then embedding $\mathbb{S}^3\hookrightarrow \mathbb{R}^4$ into Euclidean space. Under the compactification $\mathbb{R}^3\cup \{\infty\}\cong \mathbb{S}^3$, Brouwer's degree of the map $U\colon \mathbb{S}^3\to\mathbb{S}^3$ is the integral over $\mathbb{R}^3$ of the Wronskian $W_{d=3}^{k=1}$. The baryon current charge density $\mathcal{B}^0$ integrates to the baryon number (seen as Brouwer's degree): $\int_{\mathbb{R}^3} \tfrac{1}{12} \mathcal{B}^0 = \int_{\mathbb{R}^3} W_{d=3}^{k=1}$. We show that the two integrands themselves are in fact equal: the Wronskian $W_{d=3}^{k=1}$ calculates $\mathcal{B}^0$. Finally, we re-express $\mathcal{B}^0$ as both a Jacobian and a Wronskian of the pion field.
}}

\vspace{\baselineskip}

\noindent\textcolor{white!90!black}{%
\fbox{\parbox{0.975\linewidth}{%
\textcolor{white!40!black}{\begin{tabular}{lr}%
  \begin{minipage}{0.6\textwidth}%
    {\small Copyright attribution to authors. \newline
    This work is a submission to SciPost Phys. Proc. \newline
    License information to appear upon publication. \newline
    Publication information to appear upon publication.}
  \end{minipage} & \begin{minipage}{0.4\textwidth}
    {\small Received Date \newline Accepted Date \newline Published Date}%
  \end{minipage}
\end{tabular}}
}}
}




\section{Introduction}\label{SecIntro}

The Skyrme model was introduced in 1958 to describe the strong interaction~\cite{SkyrmeSkyrmions,30Anniv}. This model is a non-linear scalar field theory; it admits localised solutions (topological solitons). Each static soliton admits an integer invariant called the topological charge (or baryon number). This invariant is preserved under continuous deformations of the field. A static topological soliton is a \emph{Skyrmion} if it minimises the energy functional among the solitons with the same topological charge; see \cite{Manton87,FosterKrusch13,MantonRatmapSkyrmions,Manton}, and also \cite{Sutcliffe23} for the equivalence of Skyrme's and Yang--Mills energies.
This entire theory uses methods from differential geometry of Riemannian manifolds such as Lie groups or symmetric spaces. Formulas are written in terms of coordinates on the fields' source and target manifolds, whereby one calculates the Jacobians of the fields (seen as smooth maps) and their eigenvalues. With the present paper we aim to offer an alternative description of the setup: namely, we embed the target manifolds (with the field values) into Euclidean spaces, and we express objects such as spherical angles and area forms in the Cartesian coordinates.

Skyrmions appear also in condensed matter physics as stable vortex-like solutions of the magnetisation of atoms on a lattice~\cite{MultifacSkyrmion}; these \emph{magnetic} Skyrmions themselves form a periodic (e.g.,\ hexagonal) crystal lattice. The formula of the topological charge for the (unit) magnetisation vector $\vec{\bm}(x,y)$ in the plane
$\BBR^2\ni(x,y)$ integrates the spherical angle, yielding the number,
\begin{equation}
    \frac{1}{4\pi} \iint_{\BBR^2} \det[\vec{\bm},\partial_x \vec{\bm},\partial_y \vec{\bm}]\, dxdy \in \BBZ,
\end{equation}
that classifies the mappings $\vec{\bm}\colon \BBR^2\cup \{\infty\} \simeq \BBS^2  \to \BBS^2\subsetneq \BBR^3$; this number is Brouwer's degree of a map,\footnote{It is exactly equal to the integral $\int_{\BBS^2}\vec{\bm}^*\omega$ of the pull-back of the area form $\omega$ of the 2-sphere by the map $\vec{\bm}$ expressed in coordinates $(x,y)$ on the plane with point at infinity $\BBR^2\cup \{\infty\}$. See \S\ref{SecWronBrouwer} for further details.} or, in physics terminology, the topological charge, winding number, or baryon number (depending on context). In effect, the integrand (normalised by the area of the unit sphere) is given by the generalised Wronskian (over the base dimension $d=2$) for the three Cartesian components of the magnetisation vector $\vec{\bm}=(m^1,m^2,m^3)$. Its higher-dimensional generalisations, with $\bm\colon\BBR^d\cup\{\infty\}\simeq \BBS^d \to \BBS^d\subsetneq \BBR^{d+1}$, involve the complete generalised Wronskians $W_{d\geqslant2}^{k=1}$ of differential order $k=1$ over the $d$-dimensional bases $\BBR^d$ (under the assumption $\bm(\infty)=\bm_0$ about the asymptotic behaviour of the smooth $\BBS^d$\nobreakdash-valued field $\bm$ on $\BBR^d$). These (anti\nobreakdash-)instanton solutions of field models are explained in much detail in~\cite{MonastyrskyTGF} and~\cite{RajaramanSI}, as well as in~\cite{SchwarzTopolPhys}. Magnetic Skyrmions were an attempt to apply Skyrme's strong interaction picture in condensed matter physics. The formalism suits well to describe the skyrmion crystals with large topological numbers, see~\cite{Manton} and~\cite{MantonRatmapSkyrmions}, where polynomials and rational functions are used in the constructions. See also~\cite{FosterKrusch13} for the rational map used in the Skyrme model.

\medskip
\noindent\textbf{Wronskians as $N$-ary structures.}\quad
Lie groups and Lie algebras are fundamental in gauge field theory, in particular in the magnetic or spin systems and in many other models of (liquid) crystals and condensed matter~\cite{PolyaGFS}.
Lie algebras are deformed to homotopy Lie algebra structures, now with multi-linear anti-symmetric brackets of arities $\geqslant 2$. All these brackets are mutually constrained through the set of Jacobi-type identities. The higher Lie structures have applications in Theoretical Physics; see the recent review~\cite{Stasheff2019Review} (\S4.1 details a wide range of applications in physics) and references therein for applications, and see~\cite{HanlonWachs} for the algebraic context. Apart from the mathematical convenience of higher structures (as they serve to resolve obstructions and make proofs work), the idea of $N$-ary structures with higher arities is that they encode the collective effects beyond the pairwise interactions or beyond the factorisations into only pairwise interactions.

The generalised Wronskian determinants for functions in many variables were rediscovered many times: e.g., in Number Theory~\cite{LeVeque1956,Schmidt1980} or in the engineering sciences~\cite{Wolsson1989a}; cf.~\cite{WronskReview} for a review of Wronskians in algebraic geometry and related fields, in particular, see~\cite{MilasQSeries} for Wronskians connecting number theory and physics through the Andrews--Gordon $q$-series. Showing up not only in the above formula of topological charge, generalised Wronskians are immediately useful to detect the linear (in)dependence of differentiable functions; these determinants appear also in the proof of Roth's theorem about approximations of numbers (see~\cite{LeVeque1956,Schmidt1980}). 

Over the base dimension $d=1$, the usual Wronskian determinants of functions in one variable realise a class of homotopy deformations for Lie algebras of vector fields on the line $\BBR^1$; this was noted by Dzumadil'daev~\cite{Dzhuma2002}. 
In~\cite{ForKac}, the second author extended Dzhumadil'daev's proof to arbitrary dimension $d\geqslant 1$, thus providing a large class of (strongly) homotopy Lie algebra structures on the spaces of functions in $d$ variables; the complete generalised Wronskians $W_{d\geqslant 1}^{k\geqslant 1}$ of (highest appearing) differential order $k$ are the multi-linear anti-symmetric brackets satisfying the table of Jacobi-type identities. In the recent paper~\cite{YER25}, the second author furthered this result to the case of \textit{in}\/complete generalised Wronskians of differential orders $k>1$, thus finally allowing the bracket's arity $N$ to increase incrementally, with the minimal step $+1$ instead of the leaps between the binomial coefficients $N=\tbinom{d+k}{k}$ as it was in the earlier work~\cite{ForKac} (see also~\cite{PRG25Wronsk}). Finite\/-\/dimensional strongly homotopy Lie algebras of polynomials over $\mathbb{R}^d$, i.e.\ in any number of base variables $d\geqslant 1 $, were studied by the authors in the recent work~\cite{FinDim}, whereas their asymptotic growth -- when infinite-dimensional -- in~\cite{FiboWron}.

\centerline{\rule{3.5in}{0.7pt}}\medskip

\noindent
This text is structured as follows. First, in \S\ref{SecWronBrouwer} we show that generalised Wronskian determinants $W_{d\geqslant 1}^{k=1}$ of differential order 1 over $\BBR^d$ calculate the Brouwer degree of fields $\bm\colon\BBR^d\cup\{\infty\}\to \BBS^d$. Specifically, the integral of $W_{d\geqslant 1}^{k=1}$ yields the degree; the Wronskian itself is the pull-back of the area form 
on the target sphere $\BBS^d$ by the smooth map $\bm\colon\BBR^d\cup\{\infty\}\to\BBS^d$.
We shall study in particular the cases $d=2$ and $d=3$, that is, Skyrmions in condensed matter on the plane and in field theory (namely, the Skyrme model) over space-time, respectively. 
It is readily seen that over the base dimension $d=2$, the pull-back of the area form, i.e.\ the solid angle, is precisely the generalised Wronskian $W_{d=2}^{k=1}$. This short calculation immediately extends to arbitrary base dimension $d\geqslant 2$. Yet the standard language of the Skyrme model operates in~terms~of~the baryon~charge~density~$\mathcal{B}^0$ (see \S\ref{SecBaryonCharge}), which is apparently different~from~the~construction~of~the Wronskian $W_{d=3}^{k=1}$. We show in \S\ref{SecBaryonCharge} that they are in fact the same object \& formula!
(Over the dimension $d=3$, the field is denoted by $U$ and the target manifold is the Lie group~$SU(2)\cong \BBS^3$.)

The case $d=3$ is special in that $\BBS^3\cong SU(2)$ carries the Lie group structure such that the exponential map from the Lie algebra $\mathfrak{su}(2)$ is surjective (onto). In \S\ref{SecPionField} we express the field $U$ in terms of the coordinates in the Lie algebra (specifically, the pion field) with respect to the basis $\{i\sigma_1,i\sigma_2,i\sigma_3\}$ of Pauli matrices. Finally, we rewrite the baron charge density $\mathcal{B}^0$ via the Jacobian and via the Wronskian of the pion field. In conclusion, we reach four seemingly different formulas of the density $\mathcal{B}^0$. This paper ends with a set of open problems about bringing the unit quaternions $\BBS^3\subsetneq \mathbb{H}$ into the game as they induce the group law on $\BBS^3$.

\section{How the Wronskians calculate Brouwer's degree}\label{SecWronBrouwer}

In this section we show that by integrating the generalised Wronskian determinant of differential order $1$ over $\BBR^d\ni (x^1,\cdots , x^d)$, $W_d^{k=1}=\Bone\wedge \tfrac{\partial}{\partial x^1}\wedges \tfrac{\partial}{\partial x^d}$, where the dimension is $d\geqslant 1$, one calculates Brouwer's degree of a smooth map $\BBS^d\to\BBS^d$ between $d$\nobreakdash-spheres.

\medskip
Recall that Skyrmions exist also in condensed matter physics as vortex-like configurations of the (unit) magnetisation of atoms on a planar crystal. A magnetic material becomes a Skyrmion crystal if the vector-like configuration is tiled periodically (e.g.,\ hexagonally). By normalising the magnetisation $\vec{\bm}\colon \BBR^2\cup\{\infty\}\to \BBS^2\subsetneq \BBR^3$, the integer-valued topological invariant of a Skyrmion, the \emph{topological charge} $Q$ (i.e.\ Brouwer's degree), is given by integrating the spherical angle spanned by $\vec{\bm}$ and its partial derivative vectors:
\begin{equation}\label{EqMagneticTopolCharge}
    Q = \frac{1}{4\pi} \iint \vec{\bm}\cdot \biggl(\frac{\partial \vec{\bm}}{\partial x} \times \frac{\partial \vec{\bm}}{\partial y} \biggr)\, dx\,dy,
\end{equation}
where $4\pi=4\pi 1^2$ is the area of the unit sphere $\BBS^2$. We know that the triple product $\vec{\ba}\cdot(\vec{\boldb}\times \vec{\bc})$ is calculated by taking the determinant of the $3\times3$ matrix with rows $\vec{\ba}$, $\vec{\boldb}$, and $\vec{\bc}$. Hence
\begin{equation}\label{EqTripleProductWron}
    \vec{\bm}\cdot \biggl(\frac{\partial \vec{\bm}}{\partial x}  \times \frac{\partial \vec{\bm}}{\partial y} \biggr) = \det
    \begin{bmatrix}
        m_1 & m_2 & m_3 \\
        \partial_x m_1 & \partial_x m_2 & \partial_x m_3 \\
        \partial_y m_1 & \partial_y m_2 & \partial_y m_3
    \end{bmatrix},
\end{equation}
which is by definition (cf.\ \cite{ForKac,FinDim}) the Wronskian $W_{d=2}^{k=1}$ of differential order $1$ over $\BBR^2\ni(x,y)$ of the component functions $(m_1,m_2,m_3)=\vec{\bm}\in \BBS^2\subsetneq \BBR^3$: $\vec{\bm}\cdot \Bigl(\tfrac{\partial \vec{\bm}}{\partial x}  \times \tfrac{\partial \vec{\bm}}{\partial y} \Bigr) = W_{d=2}^{k=1}(\vec{\bm}(x,y))$. Therefore the topological charge is calculated by integrating the Wronskian $W_{d=2}^{k=1}(\vec{\bm}(x,y))$.

Let us now show that the Wronskian arises naturally in the calculation of Brouwer's degree (topological charge) of any smooth map $M\colon\BBS^2\to\BBS^2$. (The degree basically measures how many times $M$ wraps the domain sphere around the target sphere.)

\begin{prop}\label{PropBrouwerDim2}
    Let $M\colon\BBS^2\to\BBS^2$ be smooth. Under the diffeomorphism $F\colon\BBR^2\cup\{\infty\}\to \BBS^2$ given by the (inverse) stereographic projection, let $\vec{\bm}\colon\BBR^2\cup\{\infty\}\to\BBS^2$ correspond to $M$ via $M\circ F^{-1}=\vec{\bm}$. Then the Brouwer degree $Q$ of $M$ is given equivalently by
    \begin{equation}\label{EqBrouwerDim2}
        Q=\int_{\BBS^2} M^{*} \omega_{\BBS^2} = \int_{\BBR^2} \vec{\bm}^{*} \omega_{\BBS^2} = \frac{1}{\Area \BBS^2} \iint_{\BBR^2} W_{d=2}^{k=1} (\vec{\bm}(x,y))\, dx\, dy,
    \end{equation}
    where $\omega_{\BBS^2}$ is the unit-normalised area 2\nobreakdash-form on the sphere $\BBS^2$ and $*$ denotes the pull-back.
\end{prop}
\begin{proof}
    Indeed, $M$ is a representative of a class in the (Abelian) second homotopy group $\pi_2(\BBS^2)$ of the sphere, $\pi_2(\BBS^2)\cong \BBZ$. As $\pi_1(\BBS^2)\cong \{0\}$ is trivial, the group $\pi_2(\BBS^2)$ is isomorphic to the second homology group $\pi_2(\BBS^2)\cong H_2(\BBS^2)$. Taking the push-forward $M_{*}$ of $M$ that sends $2$\nobreakdash-cycles to $2$\nobreakdash-cycles $M_{*}\colon H_2(\BBS^2)\to H_2(\BBS^2)$ yields, under the isomorphism $H_2(\BBS^2)\cong \BBZ$, a group homomorphism $M_{*}\colon\BBZ\to\BBZ$. It follows that $M_{*}$ is a multiplication by some integer $Q\in\BBZ$, $M_{*}=Q\id_{H_2(\BBS^2)}$. (This integer is by definition called \textit{the Brouwer degree} of $M$.)


Let us calculate $Q$ via integration. The second de Rham cohomology of the sphere is also one-dimensional $H^2(\BBS^2;\BBR)\cong \BBR$, and has generator $[\omega]$ that satisfies $\int_{\BBS^2} \omega = 1$. Putting $\BBS^2\subsetneq \BBR^3\ni (x,y,z)$, one obtains
$\omega = \frac{1}{4\pi} (x\, dy\wedge dz - y\, dx\wedge dz + z\, dx\wedge dy)$. 
By the definition of the Riemann integral in terms of pull-backs and push-forwards, we have the identity
\begin{equation*}
    \int_{\BBS^2} M^*\omega = \int_{M_* \BBS^2} \omega = Q\int_{\BBS^2} \omega = Q
\end{equation*}
as $\omega$ is the \textit{normalised} area form. Hence the topological charge is the integral of the pull-back of the area form along the smooth map $M$.

Denote by $F$ the inverse stereographic projection\footnote{
Explicitly, $F$ is given in Cartesian coordinates $u,v$ by
$F(u,v)=(2u, 2v, r^2-1)/(r^2+1)$,
where $r^2=u^2+v^2$.
} $F\colon \BBR^2\cup \{\infty\}\to \BBS^2$. 
By definition and properties of the pull-backs of differential forms,
\begin{equation*}
    Q = \int_{\BBS^2} M^*\omega = \int_{F(\BBR^2)}M^*\omega = \int_{\BBR^2} F^*M^*\omega = \int_{\BBR^2} (M\circ F)^*\omega = \int_{\BBR^2} \vec{\bm}^*\omega,
\end{equation*}
as $\vec{\bm}(u,v)=M(F(u,v))$ by construction. Let us calculate the pull-back $\vec{\bm}^*\omega$; by definition, the pull-back along $\vec{\bm}=(m_1,m_2,m_3)$ at a point with coordinates $(u,v)\in \BBR^2$ evaluated on the two basis (tangent) vectors $\vec{e}_1=\partial/\partial u$ and $\vec{e}_2=\partial/\partial v$ of the plane $\BBR^2\ni (u,v)$ is
\begin{align*}
    \vec{\bm}^* \omega (u,v)\bigl( \tfrac{\partial}{\partial u}\wedge \tfrac{\partial}{\partial v} \bigr) &= \omega(\vec{\bm}(u,v))(\partial_u \vec{\bm}\wedge \partial_v \vec{\bm}) \\
    &= \tfrac{1}{4\pi} \bigl(m_1\,dy\wedge dz\,(\partial_u \vec{\bm}\wedge \partial_v \vec{\bm}) - m_2\, dx\wedge dz\,(\partial_u \vec{\bm}\wedge \partial_v \vec{\bm}) +\\
    &\qquad\qquad\qquad\qquad\qquad\qquad\qquad\qquad +m_3\, dx\wedge dy\,(\partial_u \vec{\bm}\wedge \partial_v \vec{\bm}) \bigr)\\
    &= \frac{1}{4\pi} \Bigl(
    m_1 \begin{vmatrix}
        \partial_u m_2 & \partial_u m_3 \\
        \partial_v m_2 & \partial_v m_3 \\
    \end{vmatrix} 
    - m_2 \begin{vmatrix}
        \partial_u m_1 & \partial_u m_3 \\
        \partial_v m_1 & \partial_v m_3 \\
    \end{vmatrix} 
    + m_3 \begin{vmatrix}
        \partial_u m_1 & \partial_u m_2 \\
        \partial_v m_1 & \partial_v m_2 \\
    \end{vmatrix} 
    \Bigr)\\
    &= \frac{1}{4\pi} \det
    \begin{vmatrix}
        m_1 & m_2 & m_3\\
        \partial_u m_1 & \partial_u m_2 & \partial_u m_3 \\
        \partial_v m_1 & \partial_v m_2 & \partial_v m_3
    \end{vmatrix}(u,v) \\
    &= \tfrac{1}{4\pi} W_{d=2}^{k=1} (m_1,m_2,m_3)(u,v) \\
    &= \tfrac{1}{4\pi} W_{d=2}^{k=1} (\vec{\bm})(u,v),
\end{align*}
i.e.\ we recover the Wronskian determinant of order one over $\BBR^2\ni (u,v)$: $W_{d=2}^{k=1}=\Bone\wedge \partial_u\wedge \partial_v$. Therefore, the expression for the topological charge becomes
\begin{equation*}
    Q = \int_{\BBS^2}M^* \omega = \int_{\BBR^2}\vec{\bm}^* \omega = \frac{1}{4\pi}\iint W_{d=2}^{k=1}(\vec{\bm})\, dxdy,
\end{equation*}
as claimed. The integrals do not diverge by construction of $\vec{\bm}$ from $M\colon\BBS^2\to \BBS^2$.
\end{proof}

As the higher homotopy groups $\pi_n(\BBS^d)$ for $n\leqslant d$ behave predictably,\footnote{The behaviour is highly non-trivial when $n>d$.} i.e.\ $\pi_{n<d}(\BBS^d)\cong\{0\}$ and $\pi_d(\BBS^d)\cong \BBZ$, the same construction of Brouwer's degree via pull-backs holds for a smooth map $M:\BBS^d\to\BBS^d$. Let us show that the expressions of the pull-backs are once again equal to the generalised Wronskian determinants, now over $\BBR^d$.

\begin{prop}\label{PropBrouwerDimAny}
    Let $M\colon \BBS^d\to \BBS^d$ ($d\geqslant 1$) be smooth. Define the map $\bm\colon \BBR^d\cup\{\infty
    \}\to \BBS^d\!\subsetneq \BBR^{d+1}$ by pre-composing $M$ with the stereographic projection $F\colon \BBR^{d}\cup\{\infty\}\to \BBS^d$: $\bm=M\circ F^{-1}$. Then Brouwer's degree $Q$ of $M$ is given by
    \begin{equation}\label{EqBrouwerDimAny}
        Q=\int_{\BBS^d} M^{*} \omega_{\BBS^d} = \int_{\BBR^d} \bm^{*} \omega_{\BBS^d} = \frac{1}{\Vol_d \BBS^2} \int_{\BBR^d} W_{d}^{k=1} (\bm(\bx))\, dx^1\cdots dx^d,
    \end{equation}
    where $\omega_{\BBS^d}$ is the unit-normalised $d$\nobreakdash-area form on $\BBS^d$ and $(x^1,\ldots,x^d)=\bx$ are the Cartesian coordinates on $\BBR^d$
\end{prop}

\begin{proof}
The constructions are also valid in $d$ dimensions as the homologies and cohomologies of the $d$-sphere $\BBS^d=\{\bx\in \BBR^{d+1}:||\bx||=1\}$ are $H_d(\BBS^d;\BBZ)\cong \BBZ$ and $H^d(\BBS^d;\BBR)\cong \BBR$. Here, the $d$-area form is the contraction of the Euclidean $(d+1)$-volume form $dx^1\wedges dx^{d+1}$ on $\BBR^{d+1}\ni(x^1,...,x^{d+1})$ by the normal vector of the sphere, obtaining
\begin{equation*}
    \omega 
    = \frac{1}{\Vol_d\BBS^d}\sum_{i=1}^{d+1}(-)^{i-1} x^i\, dx^1\wedges \widehat{dx^i}\wedges dx^{d+1}.
\end{equation*}
Again, we denote the inverse stereographic projection by $F\colon\BBR^d\cup\{\infty\}\to \BBS^d$, whereby the topological charge is $Q=\int_{\BBS^d}M^*\omega$, where $M\colon\BBS^d\to \BBS^d$ is a smooth map and the corresponding $\bm=M\circ F^{-1}=(m_1,...,m_{d+1})$ is the field over $\BBR^d\cup\{\infty\}$. Denote by $\bu=(u^1,...,u^d)$ the coordinates on $\BBR^d\cup \{\infty\}$, and define the gradient vector $\partial m_i/\partial \bu = [\partial m_i/\partial u^1,...,\partial m_i/\partial u^d]$. Then the pull-back $\bm^*\omega(\bu)\bigl(\tfrac{\partial}{\partial u^1}\wedges \tfrac{\partial}{\partial u^d}\bigr)$ of the normalised area form $\omega$ along the smooth map $\bm$ at point $\bu$, and evaluated on the $d$-blade $\partial/\partial u^1\wedges \partial/\partial u^d$ of the compactified $d$-space $\BBR^d\cup\{\infty\}$, is given by
\begin{align*}
    \frac{1}{\Vol_d \BBS^d} \sum_{i=1}^{d+1}(-)^{i-1} m_i(\bu)\cdot \det \Biggl[\frac{\partial m_1}{\partial \bu},...,\widehat{\frac{\partial m_i}{\partial \bu}},...,\frac{\partial m_{d+1}}{\partial \bu} \Biggr],
\end{align*}
where the sum is the determinant
\begin{equation*}
    \det
    \begin{bmatrix}
        m_1 & \cdots & m_{d+1} \\
        \partial m_1/\partial u^1 & \cdots & \partial m_{d+1}/\partial u^1 \\
        \vdots & \ddots & \vdots \\
        \partial m_1/\partial u^d & \cdots & \partial m_{d+1}/\partial u^d \\
    \end{bmatrix} = 
    W_d^{k=1}(m_1,...,m_{d+1})(\bu) = W_d^{k=1}(\bm)(\bu),
\end{equation*}
now the Wronskian of first order over $\BBR^d$. This again gives the equivalent expressions for the topological charge
\begin{equation*}
    Q= \int_{\BBS^d} M^*\omega = \int_{\BBR^d} \bm^*\omega = \frac{1}{\Vol_d \BBS^d} \int_{\BBR^d} W_d^{k=1}(\bm)(\bu)\, d\bu,
\end{equation*}
i.e.\ the normalised integral of the Wronskian determinant of the $d+1$ vector-component functions of $\bm$.
This completes the proof.
\end{proof}

\section[The Wronskian calculates the baryon charge density]{The Wronskian $W_{d=3}^{k=1}$ calculates the baryon charge density $\mathcal{B}^0$}\label{SecBaryonCharge}

The Skyrme model is a scalar field theory for a smooth $SU(2)$\nobreakdash-valued field $U$ over space-time: $U\colon\BBR\times\BBR^3\to SU(2)$. It is known that the model admits localised solutions, that is, $\lim_{|\bx|\to\infty}U(t;\bx)=U_\infty(t)$ in all directions, whereby one compactifies the three-dimensional space $\BBR^3$ with a point at infinity (here $U(t;\infty)=U_\infty(t)$) to the 3\nobreakdash-sphere; i.e.\ there is a diffeomorphism $\BBR^3\cup\{\infty\}\cong\BBS^3$ given by the stereographic projection. Let us emphasise that a Skyrmion is a \emph{static} localised solution: $U(t;\bx)=U(\bx)$ for all time $t$. Since $SU(2)\cong\BBS^3$ as smooth manifolds (see Definition~\ref{DefDiffeoSU}), the field $U$ is in fact a continuous map $\BBS^3\to\BBS^3$. Such maps have an integer-valued topological invariant called the Brouwer degree (of $U$), or, alternatively, the winding number, topological charge, or baryon number (specifically for the Skyrme model). 

In \S\ref{SecWronBrouwer} we showed that Brouwer's degree of a map $\BBS^d\to\BBS^d$ for any $d\in\{1,2,3,...\}$ can be calculated by integrating the generalised Wronskian determinant over $\BBR^d\cup\{\infty\}\cong \BBS^d$.
In the case of the Skyrme model one defines for a field $U$ a topological (Pontryagin\footnote{See Remark~\ref{RemBMuTopolConserved} and~\cite{Fabbri25}.}) 4\nobreakdash-current $\cB^\mu$, whose zeroth component $\cB^0$, the charge density, is constructed in such a way (see Definition~\ref{DefBaryonCurrent}) that it integrates to Brouwer's degree of the field $U$:
\begin{equation}\label{EqChargeIntegral}
    \frac{1}{\Vol_3(\BBS^3)}\int_{\BBR^3} \tfrac{1}{12}\cB^0(t;\bx)\,d^3(\bx) = \bigl(\text{Brouwer's degree of } U\bigr) = \frac{1}{\Vol_3(\BBS^3)} \int_{\BBR^3} W_{d=3}^{k=1}(\bm_U),
\end{equation}
where the second equality was proven in \S\ref{SecWronBrouwer} with $\bm_U\colon \BBR\times(\BBR^3\cup\{\infty\})\to\BBS^3$ corresponding to the field $U$ under the isomorphism $SU(2)\cong\BBS^3$.

The goal of this section is to prove that both integrands in \eqref{EqChargeIntegral} are equal: $\tfrac{1}{12}\cB^0=W_{d=3}^{k=1}(\bm)$. 

\medskip
Let us now set up the construction. Fix the spatial dimension to $d=3$, whereby spacetime becomes $\BBR\times\BBR^3$ with the Minkowski metric $(+,-,-,-)$. Let the covariant derivatives be $\partial_\mu \ceq \partial/\partial x^\mu$, with $\mu=0,1,2,3$, that is, $\partial_\bullet = \bigl(\tfrac{1}{c}   \tfrac{\partial}{\partial t}, \tfrac{\partial}{\partial x}, \tfrac{\partial}{\partial y}, \tfrac{\partial}{\partial z}\bigr)$. Throughout this section we shall work with a (not necessarily static) smooth field $U\colon \BBR\times\BBR^3\to SU(2)$. During the calculations we shall \emph{not} need to assume that the spatial component of spacetime is compactified -- we work simply over $\BBR^3$. We now recall a diffeomorphism between $SU(2)$ and $\BBS^3$.

\begin{define}\label{DefDiffeoSU}
    Embed the 3\nobreakdash-sphere into four-dimensional Euclidean space with Cartesian coordinates: $\BBS^3\hookrightarrow\BBR^4$. An element $\phi\in\BBS^3$ thus corresponds to the vector $(\phi_1,\phi_2,\phi_3,\phi_4)\in\BBR^4$, where $\phi_1^2+\cdots+\phi_4^2=1$. A matrix $U\in SU(2)$ is given by $U=\bigl(\begin{smallmatrix} \alpha & -\bar{\beta} \\ \beta & \bar{\alpha} \end{smallmatrix}\bigr)$, where $\alpha,\beta\in\BBC$ satisfy $|\alpha|^2+|\beta|^2=1$ and the bar denotes complex conjugation; upon expansion, the moduli's squares become $(\Real \alpha)^2+(\Imag \alpha)^2+(\Real \beta)^2+(\Imag \beta)^2=1$, hence there is a diffeomorphism $\Phi$,
    \begin{equation}\label{EqDiffeoSU}
        \begin{tikzcd}[/tikz/column 1/.append style={anchor=base east},
                       /tikz/column 2/.append style={anchor=base west},
                       row sep=-0.2em]
            SU(2) \arrow[r,"\Phi"] & \BBS^3\hookrightarrow\BBR^4 ,\\
            \bigl(\begin{smallmatrix} \alpha & -\bar{\beta} \\ \beta & \bar{\alpha} \end{smallmatrix}\bigr) \arrow[r,"\Phi", mapsto] & (\Real \alpha, \Imag \alpha, \Real \beta, \Imag \beta) ,\\
            \bigl(\begin{smallmatrix} 
            \phi_1+i\phi_2 & -\phi_3+i\phi_4 \\ 
            \phi_3+i\phi_4 & \phi_1-i\phi_2 
            \end{smallmatrix}\bigr) \arrow[r, "\Phi^{-1}", mapsfrom] & (\phi_1,\phi_2,\phi_3,\phi_4).
        \end{tikzcd}
    \end{equation}
    To the field $U\colon\BBR\times\BBR^3\to SU(2)$ there corresponds $\bm_U\colon\BBR\times\BBR^3\to \BBS^3$ by composing $U$ with the diffeomorphism: $\bm_U=\Phi\circ U$.\footnote{If we care about the (Lie) group structure, we must map $\bm_U:SU(2)\to \BBS^3\hookrightarrow \mathbb{H}$ (as unit quaternion subgroup) to $(\Real \alpha, \Imag \alpha, -\Real \beta, \Imag \beta)$, whereby Pauli matrices $\sigma_1,\sigma_2,\sigma_3$ correspond to the imaginary units $i,j,k\in\mathbb{H}$.}
\end{define}

\begin{define}\label{DefBaryonCurrent}
    Given a field $U\colon \BBR\times\BBR^3\to SU(2)$, define the \emph{left-current} $L_\mu=U^\dagger \partial_\mu U$ (i.e.\ $L_\mu=U^{-1}\partial_\mu U$). The (pre-normalised) \emph{baryon 4\nobreakdash-current} $\cB^\mu$ is given by the components
    \begin{equation}\label{EqBaryonCurrentBMu}
        \cB^\mu = \varepsilon^{\mu\!\nu\sigma\!\rho} \tr(L_\nu L_\sigma L_\rho),
    \end{equation}
    where all indices $\mu,\nu,\sigma,\rho$ run from $0$ to $3$ and $\varepsilon^{\mu\!\nu\sigma\!\rho}$ is the Levi--Civita symbol with $\varepsilon^{0123}=+1$. The zeroth component $\cB^0$ is called the \emph{charge density}.
\end{define}

\begin{rem}\label{RemBMuTopolConserved}
    The 4\nobreakdash-current $\cB^\mu$ is a topologically conserved current. That is, $\partial_\mu \cB^\mu=0$ for \emph{any} field $U$ (i.e.\ not necessarily a solution of Euler's equations of the Skyrme model's Lagrangian, cf.~\cite{SkyrmeSkyrmions}). Moreover, for a spatially localised field $U\colon \BBR\times (\BBR^3\cup\{\infty\})\to SU(2)$ the integral of the charge density, $\int_{\BBR^3} \cB^0(t;\bx)\, d^3\bx$, is a topological invariant. Upon the normalisation $\cB^0\mapsto \cB^0/(12\cdot 2\pi^2)$, the integral $\frac{1}{\Vol_3(\BBS^3)}\int_{\BBR^3}\tfrac{1}{12}\cB^0(t;\bx)\,d^3\bx$ becomes an integer, identified with the \emph{baryon number} of $U$.
    
    In the language of Riemannian geometry, the current $\cB^\mu$ is called a \textit{Pontryagin current}, it encodes the curvature; see~\cite{Fabbri25}, where the Pontryagin current $K^\mu=\varepsilon^{\mu\!\nu\sigma\!\rho}F_{\nu\sigma}A_\rho$ of electrodynamics is discussed (here $F_{\mu\nu}=\partial_\mu A_\nu - \partial_\nu A_\mu$ is the field tensor and $A_\mu$ is the 4\nobreakdash-potential).
\end{rem}

\begin{lem}\label{LemLeftCurrentsu2}
    The left-current $L_\mu$ belongs to the Lie algebra $\su(2)$.\footnote{
        Moreover, $L_\mu$ is a left-invariant Maurer--Cartan current for the Lie group $SU(2)$. See~\cite{30Anniv}, specifically \S2.4.
    }
\end{lem}
The proof is given in Appendix~\ref{AppLeftCurrentsu2}. As $\su(2)$ has a basis $i\sigma_1,i\sigma_2,i\sigma_3$, where $\{\sigma_a\}$ are the three Pauli matrices 
$\sigma_1=\bigl(\begin{smallmatrix} 0&1 \\ 1&0 \end{smallmatrix}\bigr)$,
$\sigma_2=\bigl(\begin{smallmatrix} 0&-i \\ i&0 \end{smallmatrix}\bigr)$, and
$\sigma_3=\bigl(\begin{smallmatrix} 1&0 \\ 0&-1 \end{smallmatrix}\bigr)$, express the $\su(2)$ current $L_\mu(t;\bx)$ in terms of this basis: $L_\mu=i\sigma_a L_\mu^a$, where the coefficients are real, $L_\mu^a\in\BBR$.

Let us now rewrite the baryon 4\nobreakdash-current charge density $\cB^0$ in terms of the coefficients $L_\mu^a$. 

\begin{prop}[\!\cite{30Anniv}, (2.4.6)]\label{PropLmatrix}
    For any field $U$ the density $\cB^0$ is given by the $3\times 3$~determinant
    \begin{equation}\label{EqBzeroLmatrix}
        \cB^0 = 12\,\det \bigl([L_i^a]_{i,a=1,2,3}\bigr),
    \end{equation}
    where the coefficient $12=2\cdot 6$ arises from calculating the trace $\tr(\sigma_a\sigma_b\sigma_c)$ and the order of the symmetry group $|S_3|=3!=6$.
\end{prop}
\begin{proof}
    Recall the identity $\sigma_a \sigma_b = \delta_{ab} \id_{2\times 2} + i\varepsilon^{abc} \sigma_c$ for Pauli matrices, and that each $\sigma_a$ is traceless. It follows that $\tr(\sigma_a\sigma_b\sigma_c)=2i\varepsilon^{abc}$, which yields the current $\cB^\mu=\varepsilon^{\mu\!\nu\sigma\!\rho}\varepsilon_{abc}L_\nu^a L_\sigma^b L_\rho^c$ by using the real coefficients $L_\mu^a$ in the $\{i\sigma_a\}$ basis of $\su(2)$. The $abc$\nobreakdash-contracted expression is $\varepsilon_{abc}L_\nu^a L_\sigma^b L_\rho^c=\det[\vec{L}_\nu, \vec{L}_\sigma, \vec{L}_\rho]$, whereby contracting with $\varepsilon^{\mu\!\nu\sigma\!\rho}=(-)^\mu \varepsilon^{\nu\sigma\!\rho}$ (where $\nu,\sigma,\rho\neq \mu$) yields cancellation of signs from every transposition, hence each cycle $(\nu\sigma\!\rho)\in S_3$ contributes with the same $(+)$ sign, in total $|S_3|=3!=6$ times, resulting in
    \begin{equation}\label{EqBmuAnyComponent}
        \cB^\mu = (-)^\mu\, 2|S_3| \det\bigl([L_i^a]_{i\neq \mu}^{a=1,2,3}\bigr).
    \end{equation}
    Set $\mu=0$ (i.e.\ $\cB^0$ is the density); then $\cB^0=12\det\bigl([L_i^a]_{i,a=1,2,3}\bigr)$, which was to be shown.
\end{proof}

We now set up two lemmas to show that the $3\times 3$ determinant \eqref{EqBzeroLmatrix} is equal to the Wronskian $W_{d=1}^{k=1}$ of $\bm_U=\Phi\circ U\colon \BBR\times \BBR^3\to \BBS^3$, where the diffeomorphism $\Phi\colon SU(2)\to \BBS^3$ is defined in~\eqref{EqDiffeoSU}.

\begin{lem}\label{LemOrthogBasis}
    Embed the target 3\nobreakdash-sphere into Euclidean space $\BBS^3\!\hookrightarrow\! \BBR^4$ and let $U\colon\BBR\times\BBR^3\to SU(2)$ be any smooth field with $\bm_U=\Phi\circ U$. Then $\hat{\be}_1,\hat{\be}_2,\hat{\be}_3$ calculated by $\hat{\be}_a=\Phi(U\cdot i\sigma_a)$ is an orthonormal basis of the tangent space $T_{\bm_U}\BBS^3$ (as an affine subspace of $\BBR^3$). Hence $\bm_U,\hat{\be}_1,\hat{\be}_2,\hat{\be}_3$ is an orthonormal basis of $\BBR^4$.
\end{lem}
\begin{proof}
    Put $\bm=(\phi_1,\phi_2,\phi_3,\phi_4)\in\BBS^3\subsetneq \BBR^4$. Then $U=\Phi^{-1}\circ \bm$ becomes $U=\bigl(\begin{smallmatrix}
        \phi_1+i\phi_2 & -\phi_3 + i\phi_4 \\
        \phi_3 + i\phi_4 & \phi_1 - i\phi_2
    \end{smallmatrix}\bigr)$.
    The calculation of the matrix products $U\cdot i\sigma_a$ yields the vectors $\hat{\be}_a\in\BBR^4$ by $\hat{\be}_a=\Phi(U\cdot i\sigma_a)$:
    \begin{align}
    \hat{\be}_1 &= 
        (-\phi_4,-\phi_3,\phi_2,\phi_1), &
    \hat{\be}_2 &= 
        (\phi_3,-\phi_4,-\phi_1,\phi_2), &&\text{and} &
    \hat{\be}_3 &= 
        (-\phi_2,\phi_1,-\phi_4,\phi_3).
    \end{align}
    Put $R=[\bm,\hat{\be}_1,\hat{\be}_2,\hat{\be}_3]$ (a $4\times 4$ matrix). A computer algebra (in \textsf{SymPy}) calculation shows that $RR^\top = |\bm|^2\id_{4\times 4}=\id_{4\times 4}$, hence $\bm,\hat{\be}_1,\hat{\be}_2,\hat{\be}_3$ is an orthonormal basis on $\BBR^4$ as claimed.
\end{proof}

\begin{lem}\label{LemWronskCoeffmatrix}
    In the notation of the previous lemma, the partial derivatives w.r.t.\ $x,y,z$ of $\bm_U$ are orthogonal to $\bm_U\in\BBS^3\hookrightarrow\BBR^4$. That is, $\partial_i\bm_U\in T_{\bm_U}\BBS^3=\Span\{\hat{\be}_1,\hat{\be}_2,\hat{\be}_3\}$. Then
    \begin{equation}\label{EqCoeffMatrixWr}
        \det\bigl([\hat{\be}_a\cdot \partial_i \bm_U]_{i,a=1,2,3}\bigr) = W_{d=3}^{k=1}(\bm_U),
    \end{equation}
    i.e.\ the determinant of the $3\times 3$ matrix of coordinates of $\partial_i\bm_U$ w.r.t.\ this basis of the tangent space is equal to the generalised Wronskian over $\BBR^3$ of $\bm_U$.
\end{lem}
\begin{proof}
    The calculation $\bm\cdot \partial_i \bm=\tfrac{1}{2}\partial_i|\bm|^2=0$ shows that $\partial_i\bm\in T_\bm\BBS^3$. Define the $4\times 4$ matrices $R=[\bm,\hat{\be}_1,\hat{\be}_2,\hat{\be}_3]$ and $S=[\bm,\partial_x\bm,\partial_y\bm,\partial_z\bm]^\top$, i.e.\ they are constructed to satisfy $\det R=1$ (by Lemma~\ref{LemOrthogBasis}) and $\det S = W_{d=3}^{k=1}(\bm)$. Their matrix product (again a $4\times 4$ matrix) is the block matrix
    \begin{equation}
        SR = \begin{bmatrix}
            \bm \\ [\partial_i\bm]
        \end{bmatrix}
        \begin{bmatrix}
            \bm & [\hat{\be}_a]
        \end{bmatrix}
        = \begin{bmatrix}
            \bm\cdot\bm & [\bm\cdot \hat{\be}_a]\\
            [\partial_i\bm\cdot\bm] & [\partial_i\bm\cdot \hat{\be}_a]
        \end{bmatrix}
        = \begin{bmatrix}
            1 & 0 \\
            0 & [\partial_i\bm\cdot \hat{\be}_a],
        \end{bmatrix}
    \end{equation}
    where the last equality follows from $\partial_i\bm \perp \bm$ and $\bm\perp \hat{\be}_a$, leaving only the $1\times 1$ unit block and the $3\times 3$ block $[\partial_i\bm\cdot \hat{\be}_a]_{i,a=1,2,3}$ on the diagonal. Taking the determinant of the product yields
    $
        1\cdot \det\bigl([\partial_i\bm\cdot \hat{\be}_a]_{i,a=1,2,3}\bigr) = \det(SR) = W_{d=3}^{k=1}(\bm)\cdot 1
    $, 
    proving the claim of the lemma~\eqref{EqCoeffMatrixWr}.
\end{proof}

Finally, comparing Eqs~\eqref{EqBzeroLmatrix} and~\eqref{EqCoeffMatrixWr}, we show that the matrix elements of both $3\times 3$ determinants are equal, whence the claim of this section follows. (See the illustration in Figure~\ref{SubfigDrawingFrame}.)

\begin{figure}[h!]
    \centering
    \begin{subfigure}{0.5\linewidth}
        \centering
\def\R{1.5}        
\def\lat{50}     
\def\lon{50}     
\def\vphi{30}    
\def\Lone{1.7}   
\def\Ltwo{1.5}   

\pgfmathsetmacro{\mx}{\R*cos(\lat)*cos(\lon)}
\pgfmathsetmacro{\my}{\R*cos(\lat)*sin(\lon)}
\pgfmathsetmacro{\mz}{\R*sin(\lat)}

\pgfmathsetmacro{\eax}{-sin(\lon)}
\pgfmathsetmacro{\eay}{ cos(\lon)}
\pgfmathsetmacro{\eaz}{0}
\pgfmathsetmacro{\ebx}{-sin(\lat)*cos(\lon)}
\pgfmathsetmacro{\eby}{-sin(\lat)*sin(\lon)}
\pgfmathsetmacro{\ebz}{ cos(\lat)}


\begin{tikzpicture}[scale=1,
  x={(-0.500cm,-0.296cm)},
  y={( 0.866cm,-0.171cm)},
  z={( 0.000cm, 0.940cm)},
  >={Stealth[length=2.6mm,width=2mm]},
  every node/.style={font=\large},
  hidden/.style={dashed, black!55, line width=0.5pt},
  front/.style={black, line width=0.9pt},
]

\begin{scope}[x={(1cm,0)}, y={(0,1cm)}]
  \draw[front] (0,0) circle[radius=\R cm];
\end{scope}

\draw[hidden, domain=\vphi+90:\vphi+270, samples=60, smooth, variable=\t]
  plot ({\R*cos(\t)}, {\R*sin(\t)}, 0);
\draw[front, domain=\vphi-90:\vphi+90, samples=60, smooth, variable=\t]
  plot ({\R*cos(\t)}, {\R*sin(\t)}, 0);

\draw[hidden, domain=\vphi+116:\vphi+244, samples=60, smooth, variable=\t]
  plot ({\R*cos(\lat)*cos(\t)}, {\R*cos(\lat)*sin(\t)}, {\R*sin(\lat)});
\draw[front, domain=\vphi-244:\vphi+116, samples=100, smooth, variable=\t]
  plot ({\R*cos(\lat)*cos(\t)}, {\R*cos(\lat)*sin(\t)}, {\R*sin(\lat)});

\foreach \a in {-180,-170,...,170}{
  \draw[line width=0.6pt, domain=\lat:90, samples=20, variable=\u]
    plot ({\R*cos(\u)*cos(\lon+\a)}, {\R*cos(\u)*sin(\lon+\a)}, {\R*sin(\u)});
}
\fill (0,0,\R) circle[radius=1.5pt];

\fill[green!50!black, opacity=0.18]
  \PP{-1.4}{-1.4} -- \PP{3}{-1.4} -- \PP{3}{2.5} -- \PP{-1.4}{2.5} -- cycle;
\draw[green!45!black, line width=0.7pt]
  \PP{-1.4}{-1.4} -- \PP{3}{-1.4} -- \PP{3}{2.5} -- \PP{-1.4}{2.5} -- cycle;

\fill[red!60!black, opacity=0.18]
  \PP{0}{0} -- \PP{\Lone}{0} -- \PP{\Lone}{\Ltwo} -- \PP{0}{\Ltwo} -- cycle;

\fill (0,0,0) circle[radius=1.5pt];
\draw[->, line width=1pt] (0,0,0) -- (\mx,\my,\mz)
  node[pos=0.55, below right=-1pt] {$\bm$};

\draw[->, line width=1pt] \PP{0}{0} -- \PP{2.5}{0}
  node[right] {$\hat{\be}_1$};
\draw[->, line width=1pt] \PP{0}{0} -- \PP{0}{2.4}
  node[above] {$\hat{\be}_2$};

\draw[red!70!black, dashed, line width=0.9pt] \PP{\Lone}{\Ltwo} -- \PP{\Lone}{0};
\draw[red!70!black, dashed, line width=0.9pt] \PP{\Lone}{\Ltwo} -- \PP{0}{\Ltwo};
\draw[->, red!70!black, line width=1.6pt] \PP{0}{0} -- \PP{\Lone}{\Ltwo}
  node[above right, red!70!black] {$\partial_i \bm$};

\draw[red!70!black, line width=2pt] \PP{0}{0} -- \PP{\Lone}{0};
\draw[red!70!black, line width=2pt] \PP{0}{0} -- \PP{0}{\Ltwo};
\node[red!70!black, below]    at \PP{\Lone}{0} {$L_i^1$};
\node[red!70!black, left] at \PP{0}{\Ltwo} {$L_i^2$};

\fill (\mx,\my,\mz) circle[radius=1.5pt];
\node[below left=-1pt] at (0,0,0) {\normalsize $O$};

\end{tikzpicture}
        \caption{}
        \label{SubfigDrawingFrame}
    \end{subfigure}%
    \begin{subfigure}{0.5\linewidth}
        \centering

\def\RT{1.5}        
\def\latT{55}     
\def\lonT{57}     
\def\vphiT{30}    
\def\elevT{20}    

\def\vxST{-0.5} \def\vxTT{ 1} \def\vxHT{2.7}   
\def\vyST{1.3} \def\vyTT{0.3} \def\vyHT{1.4}   
\def\vzST{-0.6} \def\vzTT{-1.0} \def\vzHT{1.9}   

\pgfmathsetmacro{\nxT}{cos(\latT)*cos(\lonT)}
\pgfmathsetmacro{\nyT}{cos(\latT)*sin(\lonT)}
\pgfmathsetmacro{\nzT}{sin(\latT)}
\pgfmathsetmacro{\mxT}{\RT*\nxT}
\pgfmathsetmacro{\myT}{\RT*\nyT}
\pgfmathsetmacro{\mzT}{\RT*\nzT}
\pgfmathsetmacro{\eaxT}{-sin(\lonT)}
\pgfmathsetmacro{\eayT}{ cos(\lonT)}
\pgfmathsetmacro{\eazT}{0}
\pgfmathsetmacro{\ebxT}{ sin(\latT)*cos(\lonT)}
\pgfmathsetmacro{\ebyT}{ sin(\latT)*sin(\lonT)}
\pgfmathsetmacro{\ebzT}{-cos(\latT)}

\pgfmathsetmacro{\dlatT}{acos(-tan(\latT)*tan(\elevT))}
\pgfmathsetmacro{\lminT}{-atan(cos(\elevT)*cos(\lonT-\vphiT)/sin(\elevT))}

\begin{tikzpicture}[scale=1,
  x={(-0.500cm,-0.296cm)},
  y={( 0.866cm,-0.171cm)},
  z={( 0.000cm, 0.940cm)},
  >={Stealth[length=2.6mm,width=2mm]},
  every node/.style={font=\large},
  hidden/.style={dashed, black!55, line width=0.5pt},
  front/.style={black, line width=0.9pt},
]

\begin{scope}[x={(1cm,0)}, y={(0,1cm)}]
  \draw[front] (0,0) circle[radius=\RT cm];
\end{scope}

\draw[hidden, domain=\vphiT+90:\vphiT+270, samples=60, smooth, variable=\t]
  plot ({\RT*cos(\t)}, {\RT*sin(\t)}, 0);
\draw[front, domain=\vphiT-90:\vphiT+90, samples=60, smooth, variable=\t]
  plot ({\RT*cos(\t)}, {\RT*sin(\t)}, 0);

\draw[hidden, domain=\vphiT+\dlatT:\vphiT+360-\dlatT, samples=60, smooth, variable=\t]
  plot ({\RT*cos(\latT)*cos(\t)}, {\RT*cos(\latT)*sin(\t)}, {\RT*sin(\latT)});
\draw[front, domain=\vphiT-\dlatT:\vphiT+\dlatT, samples=100, smooth, variable=\t]
  plot ({\RT*cos(\latT)*cos(\t)}, {\RT*cos(\latT)*sin(\t)}, {\RT*sin(\latT)});

\foreach \a in {-180,-170,...,170}{
  \draw[line width=0.6pt, domain=\latT:90, samples=20, variable=\u]
    plot ({\RT*cos(\u)*cos(\lonT+\a)}, {\RT*cos(\u)*sin(\lonT+\a)}, {\RT*sin(\u)});
}
\fill (0,0,\RT) circle[radius=1.5pt];


\fill (0,0,0) circle[radius=2.5pt];
\node[below left=-1pt] at (0,0,0) {\normalsize $O$};
\draw[->, line width=1pt] (0,0,0) -- (\mxT,\myT,\mzT)
  node[pos=0.5, left] {$\hat{\btheta}$};

\fill[green!50!black, opacity=0.18]
  \PPT{-2.5}{-2.1} -- \PPT{3.0}{-2.1} -- \PPT{3.0}{3.0} -- \PPT{-2.5}{3.0} -- cycle;
\draw[green!45!black, line width=0.8pt]
  \PPT{-2.5}{-2.1} -- \PPT{3.0}{-2.1} -- \PPT{3.0}{3.0} -- \PPT{-2.5}{3.0} -- cycle;

\draw[->, line width=1pt] \PPT{0}{0} -- \PPT{2.4}{0}
  node[right] {$\hat{\bepsilon}_2$};
\draw[->, line width=1pt] \PPT{0}{0} -- \PPT{0}{2.4}
  node[below] {$\hat{\bepsilon}_1$};

\fill[blue!60, opacity=0.30]
  \PPT{0}{0} -- \QQT{\vxST}{\vxTT}{\vxHT} -- \QQT{\vyST}{\vyTT}{\vyHT} -- cycle;
\fill[blue!60, opacity=0.30]
  \PPT{0}{0} -- \QQT{\vyST}{\vyTT}{\vyHT} -- \QQT{\vzST}{\vzTT}{\vzHT} -- cycle;
\draw[blue!60!black, line width=0.6pt] \PPT{0}{0} -- \QQT{\vyST}{\vyTT}{\vyHT};
\fill[blue!60, opacity=0.30]
  \QQT{\vxST}{\vxTT}{\vxHT} -- \QQT{\vyST}{\vyTT}{\vyHT} -- \QQT{\vzST}{\vzTT}{\vzHT} -- cycle;
\fill[blue!70, opacity=0.30]
  \PPT{0}{0} -- \QQT{\vxST}{\vxTT}{\vxHT} -- \QQT{\vzST}{\vzTT}{\vzHT} -- cycle;
\draw[blue!60!black, line width=0.7pt]
  \QQT{\vxST}{\vxTT}{\vxHT} -- \QQT{\vyST}{\vyTT}{\vyHT} -- \QQT{\vzST}{\vzTT}{\vzHT} -- cycle;
\draw[blue!60!black, line width=0.7pt] \PPT{0}{0} -- \QQT{\vxST}{\vxTT}{\vxHT};
\draw[blue!60!black, line width=0.7pt] \PPT{0}{0} -- \QQT{\vzST}{\vzTT}{\vzHT};

\draw[->, line width=1pt, green!45!black, line width=1.5pt] \PPT{0}{0} -- \QQT{0}{0}{\RT}
  node[above right=-4pt] {$\hat{\btheta}$}; 

\draw[->, red!70!black, line width=1.5pt] \PPT{0}{0} -- \PPT{\vxST}{\vxTT}
  node[below right=-6pt]       {$\partial_x\hat{\btheta}$};

\draw[->, red!70!black, line width=1.5pt] \PPT{0}{0} -- \PPT{\vyST}{\vyTT}
  node[below right=-5pt] {$\partial_y\hat{\btheta}$};
\draw[->, red!70!black, line width=1.5pt] \PPT{0}{0} -- \PPT{\vzST}{\vzTT}
  node[left=-2pt]        {$\partial_z\hat{\btheta}$};

\draw[->, line width=0.9pt] \PPT{\vyST}{\vyTT} -- \QQT{\vyST}{\vyTT}{\vyHT}
  node[above right=-4pt] {$\textcolor{red!70!black}{\partial_y \hat{\btheta}}+(\partial_y\theta)\textcolor{green!45!black}{\hat{\btheta}}$};
\draw[->, line width=0.9pt] \PPT{\vzST}{\vzTT} -- \QQT{\vzST}{\vzTT}{\vzHT}
  node[above]       {$\textcolor{red!70!black}{\partial_z \hat{\btheta}}+(\partial_z\theta)\textcolor{green!45!black}{\hat{\btheta}}$};

\draw[->, line width=0.9pt] \PPT{\vxST}{\vxTT} -- \QQT{\vxST}{\vxTT}{\vxHT};
\draw[->, line width=0.9pt] \PPT{\vyST}{\vyTT} -- \QQT{\vyST}{\vyTT}{\vyHT};
\draw[->, line width=0.9pt] \PPT{\vzST}{\vzTT} -- \QQT{\vzST}{\vzTT}{\vzHT};

\fill (\mxT,\myT,\mzT) circle[radius=1.5pt];
\fill \QQT{\vxST}{\vxTT}{\vxHT} circle[radius=1.5pt];
\fill \QQT{\vyST}{\vyTT}{\vyHT} circle[radius=1.5pt];
\fill \QQT{\vzST}{\vzTT}{\vzHT} circle[radius=1.5pt];
\fill \QQT{\vxST}{\vxTT}{0} circle[radius=1.5pt];
\fill \QQT{\vyST}{\vyTT}{0} circle[radius=1.5pt];
\fill \QQT{\vzST}{\vzTT}{0} circle[radius=1.5pt];

\end{tikzpicture}
        \caption{}
        \label{SubfigDrawingTetra}
    \end{subfigure}
    \caption{(\textit{\subref{SubfigDrawingFrame}})  An illustration of the relation between the tangent vector $\partial_i \bm \in T_{\bm} \BBS^3=\Span\{\hat{\be}_1,\hat{\be}_2,\hat{\be}_3\}$ and the column vector $\vec{L}_i=(L_i^1,L_i^2,L_i^3)$ of coefficients $L_i^a$ (i.e.\ $L_i=i\sigma_a L_i^a\in \su(2)$). Here the illustration is made on the 2\nobreakdash-sphere $\BBS^2\ni \text{``}\bm\text{''}$ instead of the 3\nobreakdash-sphere (whereby $\vec{L}_i=(L_i^1,L_i^2)\in \BBR^2$) as the authors do not possess four-dimensional sight. See Lemma~\ref{LemOrthogBasis} and the proof of Proposition~\ref{PropBzeroWronskian}.\\
    \phantom{}\quad(\textit{\subref{SubfigDrawingTetra}}) Geometric interpretation of the determinant quantity, Eq.~\eqref{EqDetLProvisional}, from Lemma~\ref{LemCoordVectorL}. The quantity from~\eqref{EqDetLProvisional} is equal to the volume of the tetrahedron in $T_{\hat{\btheta}}\BBS^2\oplus\textcolor{green!45!black}{\hat{\btheta}}\cong \BBR^3$ with one 
    vertex at the origin of $T_{\hat{\btheta}}\BBS^2$ and others at the translated points $\textcolor{red!70!black}{\partial_i \hat{\btheta}}+(\partial_i\theta)\textcolor{green!45!black}{\hat{\btheta}}$.
    }
    \label{FigTwoDrawings}
\end{figure}

\begin{prop}\label{PropBzeroWronskian}
    Following the notation of Lemmas~\ref{LemOrthogBasis} and~\ref{LemWronskCoeffmatrix}, the (real) coefficients $L_i^a$ of $L_i$ in the basis $\{i\sigma_a\}$ of Pauli matrices equal the coordinates of $\partial_i\bm_U$ in the orthonormal basis $\hat{\be}_1,\hat{\be}_2,\hat{\be}_3$ of $T_{\bm_U}\BBS^3\subsetneq \BBR^4$: $L_i^a=\hat{\be}_a\cdot \partial_i\bm_U$, where the dot product is the standard Euclidean product on $\BBR^4$. Therefore the baryon 4\nobreakdash-current charge density $\cB^0$ of the field $U\colon\BBR\times\BBR^3\to SU(2)$ is equal to the generalised Wronskian over $\BBR^3$:
    \begin{equation}\label{EqBzeroWronsk}
        \tfrac{1}{12}\cB^0 = W_{d=3}^{k=1}(\bm_U),
    \end{equation}
    where $\bm_U=\Phi\circ U$ using the isomorphism $\Phi$ from~\eqref{EqDiffeoSU}.
\end{prop}
\begin{proof}
    Let us set up an isometry between the two vector spaces in which $\{\hat{\be}_a\}$ and $\{\sigma_a\}$ reside: that is, between $\BBR^4$ and the vector space of Hermitian matrices, on which the inner products are $\bv\cdot\bw=\delta_{ij}v^iw^j$, the Euclidean product on $\BBR^4$, and $\langle A,B\rangle = \tfrac{1}{2}\tr(A^\dagger B)$, respectively. Define the isometry map by
    \begin{equation}
        \begin{tikzcd}[/tikz/column 1/.append style={anchor=base east},
                       /tikz/column 2/.append style={anchor=base west},
                       every arrow/.append style={shift left},
                       row sep=-0.2em]
            (\BBR^4,\cdot) \arrow[r,"\sim"] & \Bigl(\bigl\{\bigl(
            \begin{smallmatrix}
                \alpha & -\bar{\beta} \\ \beta & \bar{\alpha}
            \end{smallmatrix}\bigr) : \alpha,\beta\in\BBC\bigr\}, \langle\cdot,\cdot\rangle \Bigr) \\
            \bv_A\ceq (a_1,a_2,b_1,b_2) \arrow[r,mapsto] & \bigl(
            \begin{smallmatrix}
                a_1+ia_2 & -b_1+ib_2 \\ b_1+ib_2 & a_1-ia_2
            \end{smallmatrix}
            \bigr) \mathrel{{=}{:}} A.
        \end{tikzcd}
    \end{equation}
    Let us now show that this mapping preserves the inner products (i.e.\ it is an isometry -- isomorphism of inner product spaces):
    \begin{align*}
        \bv_A\cdot\bv_B &= a_1c_1 + a_2c_2 + b_1d_1 + b_2d_2,\quad \text{where}\quad A=\bigl(\begin{smallmatrix}
            a & -\bar{b} \\ b & \bar{a}
        \end{smallmatrix}\bigr)\quad \text{and}\quad B=\bigl(\begin{smallmatrix}
            c & -\bar{d} \\ d & \bar{c}
        \end{smallmatrix}\bigr),\\
        \langle A,B\rangle &= \tfrac{1}{2}(\bar{a}c+\bar{b}d+\overline{\bar{a}c+\bar{b}d}) = \tfrac{1}{2}\,2\Real(\bar{a}c+\bar{b}d) = \bv_A\cdot \bv_B
    \end{align*}
    by direct expansion of the final equality.
    It remains to show that $\hat{\be}_a\cdot \partial_i \bm_U = L_i^a$, where $L_i=i\sigma_aL_i^a=U^\dagger \partial_i U$. Put $A=\Phi^{-1}(\hat{\be}_a)=Ui\sigma_a$ and $B=\Phi^{-1}(\partial_i\bm_U)=\partial_i U$. The isometry~yields
    \begin{equation*}
        \hat{\be}_a\cdot \partial_i\bm_U = \tfrac{1}{2}\tr [(Ui\sigma_a)^\dagger \partial_i U)] = \tfrac{1}{2}\tr[\sigma_a(-i)U^\dagger\partial_i U] = \tfrac{1}{2}\tr[\sigma_a(-i)i\sigma_bL_i^b]=\tfrac{1}{2}\,2\delta_{ab}L_i^b=L_i^a,
    \end{equation*}
    completing the proof.
\end{proof}

We have shown that for any smooth $SU(2)$\nobreakdash-valued field $U=\bigl(\begin{smallmatrix} \alpha & -\bar{\beta} \\ \beta & \bar{\alpha} \end{smallmatrix}\bigr)$, the baryon 4\nobreakdash-current charge density $\cB^0$ of $U$ defined by~\eqref{EqBaryonCurrentBMu} is equivalently given by the three formulas
\begin{equation}
    \tfrac{1}{12}\cB^0 = \tfrac{1}{2}\varepsilon^{\nu\sigma\!\rho} a_\rho b_\nu \bar{b}_\sigma = W_{d=3}^{k=1}(\bm_U),
\end{equation}
where the first equality, with $a_\rho=\bar{\alpha}\partial_\rho\alpha + \bar{\beta}\partial_\rho \beta\in i\BBR\subsetneq\BBC$ and $b_\nu=\alpha \partial_\nu \beta - \beta \partial_\nu \alpha \in \BBC$, was proven in Appendix~\ref{AppLeftCurrentsu2} and the second -- in Proposition~\ref{PropBzeroWronskian}.

\section[Jacobians and Wronskians: charge density in terms of the pion field]{Jacobians and Wronskians: $\mathcal{B}^0$ in terms of the pion field}\label{SecPionField}

So far we have never used the fact that the Lie group $SU(2)$ has the Lie algebra $\su(2)$ with a \emph{surjective} (onto) exponential map $\exp:\su(2)\to SU(2)$ (given by the matrix exponential). Let us advance now by using this property of the target space in Skyrme's model. 

By default, we refer to \S\ref{SecBaryonCharge} for notation and constructions. Write the field $U\colon \BBR\times \BBR^3\to SU(2)$ as $U(t;\bx)=\exp\bigl(i\vec{\sigma}\cdot \vec{\theta}(t;\bx)\bigr)$, where $\vec{\sigma}=(\sigma_1,\sigma_2,\sigma_3)$ is the vector consisting of Pauli matrices and $\vec{\theta}\colon\BBR\times\BBR^3\to\BBR^3$ is the coordinate vector $\vec{\theta}=(\theta_1,\theta_2,\theta_3)$ of an element in the Lie algebra $\su(2)$ w.r.t.\ the basis $\{i\sigma_a\}$. The field $\vec{\theta}$ is called the \emph{pion field}.\footnote{The pion field is sometimes alternatively defined and denoted by $\vec{\pi}=(\pi_1,\pi_2,\pi_3)$ such that the $SU(2)$ field $U$ becomes $U=\bigl(\begin{smallmatrix}
    \sigma+i\pi_3 & i\pi_1+\pi_2 \\ i\pi_1-\pi_2 & \sigma - i\pi_3
\end{smallmatrix}\bigr)$, where the scalar field $\sigma$ imposes $\sigma^2+|\vec{\pi}|^2=1$. That is, $(\sigma,\vec{\pi})\in\BBS^3\subsetneq\BBR^4$. By the final result of Lemma~\ref{LemUMatrixExp}, the correspondence between $(\sigma,\vec{\pi})$ and $\vec{\theta}$ is $\sigma=\cos\theta$ and $\vec{\pi}=\sin\theta\,\hat{\bn}$.
}

\begin{lem}\label{LemUMatrixExp}
    Let $U$ be an $SU(2)$\nobreakdash-valued field. The series expansion of  
    $U=\exp(i\vec{\sigma}\cdot\vec{\theta})$ yields
    \begin{equation}\label{EqFieldCosSin}
        U=\cos\theta\, \id_{2\times 2} + i\vec{\sigma}\cdot\vec{\theta}\, \sin\theta,
    \end{equation}
    where $\theta\ceq |\vec{\theta}|$. Therefore, the corresponding mapping $\bm_U=\Phi\circ U$, where the isomorphism $\Phi\colon SU(2)\to\BBS^3$ was defined  in~\eqref{EqDiffeoSU}, is given by
    \begin{equation}\label{EqmUCosSin}
        \bm_U = \bigl(\cos\theta,\tfrac{\sin\theta}{\theta}(\theta_3,-\theta_2,\theta_1)\bigr) = (\cos\theta,\sin\theta\,\hat{\bn}),
    \end{equation}
    where $\hat{\bn}\ceq(\theta_3/\theta,-\theta_2/\theta,\theta_1/\theta)$ is a unit vector, $\hat{\bn}\in\BBS^3\hookrightarrow\BBR^3$.
\end{lem}
\begin{proof}
    Indeed, $(\vec{\sigma}\cdot\vec{\theta})^2=\theta^2\id_{2}$, whence $(\vec{\sigma}\cdot\vec{\theta})^{2k}=(-)^k\theta^{2k}\id_{2}$ and $(\vec{\sigma}\cdot\vec{\theta})^{2k+1}=(-)^k\theta^{2k} i\vec{\sigma}\cdot\vec{\theta}$ follow. Dividing each term by the respective factorial, i.e.\ $(2k)!$ and $(2k+1)!$, yields the series expansions of $\cos\theta$ and $\sin\theta$, respectively, proving~\eqref{EqFieldCosSin}. Expanding the product $i\vec{\sigma}\cdot\vec{\theta}\sin\theta$ and applying the diffeomorphism $\Phi$ (cf.~\eqref{EqDiffeoSU}) to $U$ gives~\eqref{EqmUCosSin}, completing the proof.
\end{proof}

Recall that the baryon 4\nobreakdash-current charge density $\cB^0$ is given by the $3\times 3$ determinant $\tfrac{1}{12}\cB^0=\det\bigl([L_i^a]_{i,a=1,2,3}\bigr)$ in terms of the coordinates $(L_i^1,L_i^2,L_i^3)\mathrel{{=}{:}}\vec{L}_i$ of the left-$\su(2)$\nobreakdash-current $L_i$ in the $\{i\sigma_a\}$ basis of $\su(2)$: $L_i=i\sigma_aL_i^a$. (Cf.\ Lemma~\ref{LemLeftCurrentsu2}, Proposition~\ref{PropLmatrix}, and (2.4.6) in~\cite{30Anniv}.)

Let us now write the coordinate vectors $\vec{L}_i$ in terms of the pion field $\vec{\theta}$ by using the equality $L_i^a=\hat{\be}_a\cdot \partial_i\bm_U$ from Proposition~\ref{PropBzeroWronskian}. We then express the $3\times3$ determinant in terms of $\theta=|\vec{\theta}|$ and $\hat{\btheta}=\vec{\theta}/\theta$ in a provisional way; we shall then immediately see that the resulting expression is equal to two familiar determinants from calculus and differential geometry: the Jacobian determinant and the generalised Wronskian determinant.

\begin{lem}\label{LemCoordVectorL}
    The coordinate vector $\vec{L}_i\in\BBR^3$ of $L_i\in\su(2)=\Span\{i\sigma_1,i\sigma_2,i\sigma_3\}$ is given by
    \begin{equation}\label{EqVectorL}
        \vec{L}_i=\hat{\btheta}\partial_i\theta + \sin\theta\, R_\theta(\partial_i\hat{\btheta}),
    \end{equation}
    where $R_\theta$ is a rotation operator ($2\times 2$ matrix) by angle $\theta=|\vec{\theta}|$ in the tangent plane $T_{\hat{\btheta}}\BBS^2\subsetneq \BBR^3$ (embedded as a true vector subspace). Hence the $3\times 3$ determinant becomes
    \begin{equation}\label{EqDetLProvisional}
        \det\bigl([L_i^a]_{i,a=1,2,3}\bigr) = \sin^2\theta\, \varepsilon^{i\!jk}\partial_i\theta\, p_j\, q_k,
    \end{equation}
    where $p_i$ and $q_i$ are the coordinates of $\partial_i\hat{\btheta}$ w.r.t.\ a right-handed orthonormal frame $\hat{\btheta},\hat{\boldsymbol{\epsilon}}_1,\hat{\boldsymbol{\epsilon}}_2$ of $\BBR^3$ (i.e.\ $\Span\{\hat{\boldsymbol{\epsilon}}_1,\hat{\boldsymbol{\epsilon}}_2\}=T_{\hat{\btheta}}\BBS^2$).
\end{lem}

The proof is given in Appendix~\ref{AppProofLemmaL}. See the geometric interpretation of the l.\!\nobreakdash-h.s. of~\eqref{EqDetLProvisional} in Figure~\ref{SubfigDrawingTetra} on p.~\pageref{SubfigDrawingTetra}; that is, $\varepsilon^{i\!jk}\partial_i\theta\, p_j\, q_k = 6\cdot \Vol_3(\text{tetrahedron})$. (The interpretation of the coefficients on the r.\!\nobreakdash-h.s. is given in~Figure~\ref{SubfigDrawingFrame}.)

\begin{prop}\label{PropWronJacobiPion}
    The determinant~\eqref{EqDetLProvisional}, and hence baryon charge density $\cB^0$, is equal to (\textit{i})~the generalised Wronskian over $\BBR^3$:
    \begin{equation}\label{EqBzeroWronPion}
        \tfrac{1}{12}\cB^0 = \det\bigl([L_i^a]_{i,a=1,2,3}\bigr) = -\sin^2\theta\, W_{d=3}^{k=1}(\theta,\vec{\theta}/\theta),
    \end{equation}
    i.e.\ the density $\cB^0$ is once again a Wronskian determinant, this time of the pion field (cf.\ Proposition~\ref{PropBzeroWronskian}); and 
    (\textit{ii})~the Jacobian of the pion field at a fixed time, $\vec{\theta}\colon\BBR^3\to\BBR^3$:
    \begin{equation}\label{EqBzeroJacobiPion}
        \tfrac{1}{12}\cB^0 = \frac{\sin^2\theta}{\theta^2} \left|\frac{\partial{\vec{\theta}}}{\partial\bx}\right|,
    \end{equation}
    where $(\partial\vec{\theta}/\partial\bx)_{ij}=\partial \theta_i/\partial x^j$ and $i,j=1,2,3$.
\end{prop}
\begin{proof}
    (\textit{i}) Recall from the above lemma that $\hat{\btheta},\hat{\bepsilon}_1,\hat{\bepsilon}_2$ is a positive ($\hat{\btheta}\times\hat{\bepsilon}_1=\hat{\bepsilon}_2$) orthonormal basis on $\BBR^3$ and $\partial_i\hat{\btheta}=p_i\hat{\bepsilon}_1+q_i\hat{\bepsilon}_2$ in terms of the tangent space basis -- cf.~\eqref{EqDetLProvisional}. Note that $\hat{\btheta}\cdot (\partial_j\hat{\btheta}\times \partial_k \hat{\btheta})=p_jq_k-p_kq_j$, which we insert in~\eqref{EqDetLProvisional}, obtaining
    \begin{equation}\label{EqSemiWronsk}
        \det\bigl([L_i^a]\bigr) = \tfrac{1}{2} \sin^2\theta\, \varepsilon^{i\!jk}\partial_i\theta W_{x^jx^k}(\hat{\btheta})=\sin^2\theta \bigl(\partial_x\theta\, W_{yz}(\hat{\btheta}) - \partial_y\theta\, W_{xz}(\hat{\btheta}) + \partial_z\theta\, W_{xy}(\hat{\btheta}) \bigr),
    \end{equation}
    where a factor $\tfrac{1}{2}$ appears due to double counting transpositions $j\leftrightarrows k$ and $W_{x^jx^k}=\Bone\wedge \tfrac{\partial}{\partial x^j}\wedge \tfrac{\partial}{\partial x^k}$ is a generalised Wronskian in variables $x^j$ and $x^k$, which upon contraction with $\varepsilon^{i\!jk}$ yields a coefficient 2, cancelling the factor $\tfrac{1}{2}$. Expression~\eqref{EqSemiWronsk} reduces to the block-matrix determinant
    \begin{equation}\label{EqWronWithZero}
        \det\bigl([L_i^a]\bigr) = -\sin^2\theta \det \begin{bmatrix}
            0 & \hat{\btheta} \\
            [\partial_i \theta] & \partial \hat{\btheta}/\partial \bx 
        \end{bmatrix}.
    \end{equation}
    The Jacobian $\partial \hat{\btheta}/\partial \bx$ of the unit vector field $\hat{\btheta}$ appearing in~\eqref{EqWronWithZero} is, of course, singular. Hence, by the expansion of the determinant~\eqref{EqWronWithZero}, any value in the top-left corner yields the same result. Replacing the $0$ in the top-left corner of~\eqref{EqWronWithZero} by $\theta$ gives the $4\times 4$ Wronskian of $(\theta,\hat{\btheta})$.

    (\textit{ii}) As $\vec{\theta}=\theta \hat{\btheta}$, the partial derivatives $\partial_j \hat{\btheta}$ become $\partial_j \hat{\btheta} = (\partial_j\theta)\hat{\btheta} + \theta \partial_j \hat{\btheta}$, and combining this with $\partial_j \hat{\btheta} = p_j \hat{\bepsilon}_1 + q_j\hat{\bepsilon}_2$ yields $\partial_j \vec{\theta} = [\partial_j \theta, \theta p_j, \theta q_j]$ as the column vector in the basis $\hat{\btheta}, \hat{\bepsilon}_1, \hat{\bepsilon}_2$ of $\BBR^3\ni \partial_j \vec{\theta}$. Therefore the Jacobian becomes
    \begin{equation}
        \left|\frac{\partial \vec{\theta}}{\partial \bx}\right| = \theta^2 \cdot \tfrac{1}{\sin^2\theta} \det\bigl([L_i^a]\bigr),
    \end{equation}
    where the second equality follows from taking out $\theta$ from rows 2 and 3, obtaining the determinant~\eqref{EqDetLProvisional}. It follows that $\tfrac{1}{12}\cB^0 = \det\bigl([L_i^a]\bigr) = \tfrac{\sin^2\theta}{\theta^2} |\partial \vec{\theta}/\partial \bx|$ as claimed.
\end{proof}

At this point we have at our disposal four new formulas of the baryon 4\nobreakdash-current charge density $\cB^0$ of an $SU(2)$\nobreakdash-valued field $U$, defined by the trace in~\eqref{EqBaryonCurrentBMu}. Namely,
\begin{equation}\label{EqFourNewFormulas}
    \tfrac{1}{12}\cB^0 = \tfrac{1}{2}\varepsilon^{\nu\sigma\!\rho}a_\rho b_\nu \bar{b}_\sigma = W_{d=3}^{k=1}(\bm_U) = -\sin^2\theta\, W_{d=3}^{k=1}(\theta,\vec{\theta}/\theta)=\frac{\sin^2\theta}{\theta^2} \left|\frac{\partial{\vec{\theta}}}{\partial\bx}\right|,
\end{equation}
where the first two formulas were found in \S\ref{SecBaryonCharge} and the last two -- in this section.

\medskip
There is a way to construct a field of almost minimal energy (a Skyrmion field has minimal energy) by using rational holomorphic functions (\cite{MantonRatmapSkyrmions}, see also~\cite{Manton,FosterKrusch13,Sutcliffe23}). One then does not have to integrate: the topological charge is equal to the maximum of polynomial degrees in the numerator vs denominator of the rational function~\cite{FosterKrusch13}.

But suppose now that a Skyrmion is obtained in a way not referring to rational holomorphic functions from that specific construction. If so, it may still be necessary to integrate the baryon charge density $\cB^0$ in order to know the topological charge (i.e.\ the baryon number). In addition, the charge density itself encodes far more information than just the charge (i.e.\ a single number), and is used -- see~\cite{FosterKrusch13} Our new formulas~\eqref{EqFourNewFormulas} of $\cB^0$ could simplify this calculation.

\section{Conclusion}

We re-expressed a basic element of the Skyrme model with field $U\colon \BBR^3\cup\{\infty\}\to SU(2)\cong \BBS^3$, namely, the baryon current charge density $\mathcal{B}^0$, in terms of the generalised Wronskian $W_{d=3}^{k=1}$ over the multi-dimensional base -- the space $\BBR^3$ component of space-time $\BBR\times\BBR^3$. By this, we confirm that generalised Wronskians are natural in spherical \& Euclidean geometries: they calculate the pull-backs of area $d$\nobreakdash-forms on $d$\nobreakdash-spheres, whenever $\BBS^3\hookrightarrow\BBR^{d+1}$ are embedded into Euclidean spaces and Cartesian coordinates are used on $\BBR^{d+1}$. 
The case of the Skyrme model fixes the dimension $d=3$, where the target sphere $\BBS^3$ acquires a Lie group structure.
Yet our approach, with formulas given by the generalised Wronskian determinants, extends to $\sigma$\nobreakdash-models over arbitrary base dimensions $d\geqslant 1$, i.e.\ with Langrangians defined over $\BBR^d$ and admitting soliton solutions.

\subsection*{Discussion}

Finally, let us comment on the result actually proven in \S\ref{SecBaryonCharge}, and let us state two open problems and outline a wider perspective for future research.

\begin{prop}
    In the notation of \S\ref{SecBaryonCharge} (where $U\colon \BBR\times \BBR^3 \to SU(2)\cong \BBS^3$ so that the through-map was $\bm_U\colon \BBR\times\BBR^3\to\BBS^3$, whence $\mathcal{B}^\mu$ was defined by formula~\eqref{EqBaryonCurrentBMu}), every component $\mathcal{B}^\mu$ (of the four, as $\mu=0,1,2,3$) is given by the Wronskian determinant of size $4\times 4$:
    \begin{equation}
        \mathcal{B}^\mu = 12(-)^\mu\, \det \bigl[\bm_U, \partial_0\bm_U, \ldots , \widehat{\partial_\mu \bm_U},\ldots, \partial_3 \bm_U \bigr],
    \end{equation}
    where the hat denotes the term which is omitted.
\end{prop}

Indeed, by tracing through the construction in \S\ref{SecBaryonCharge}, we notice that the calculation of $\mathcal{B}^k$ ($k=1,2,3$) but not $\mathcal{B}^0$ would only require relabelling the variables; the appearance of the additional sign $(-)^k$ is due to the conversion of the 4\nobreakdash-indexed Levi--Civita symbol $\varepsilon^{\mu\!\nu\sigma\!\rho}$ into the 3\nobreakdash-indexed symbol $\varepsilon^{\nu\sigma\!\rho}$ with indices running over the three element set $\{0,\ldots,\widehat{\mu},\ldots,3\}$. The compactification was needed only near the end of the reasoning, namely, for the computation of the topological charge from its density. See Eq.~\eqref{EqBmuAnyComponent} on p.~\pageref{EqBmuAnyComponent}.

\medskip
Let us now state two open problems.

\begin{open}\label{OpenIso}
    Recall that the 3\nobreakdash-sphere $\BBS^3$, viewed as the subset $\BBS^3\subsetneq \mathbb{H}$ of unit quaternions, inherits the quaternions' group structure. However, we notice that the diffeomorphism in \S\ref{SecBaryonCharge} between the manifolds $\BBS^3$ and $SU(2)$ was not a group isomorphism (w.r.t.\ to the above group law on $\BBS^3$). The problem is to express the baryon charge density $\mathcal{B}^0$ as a Wronskian but using the Lie group isomorphism.\footnote{We expect that the only change in formulas would amount to an extra minus sign in front of the Wronskian.}
\end{open}

\begin{open}\label{OpenC2}
    Denote by $\bm$ the map from $\BBR^3\ni(x,y,z)$ to the complex space $\BBC^2\ni(\alpha,\beta)$. Having identified $\BBC^2\cong\BBR^4$ as manifolds in \S\ref{SecBaryonCharge}, we wrote formula~\eqref{EqBzeroWronsk} with the Wronskian of $\bm$. The problem is to develop a good definition of the Wronskian into the target $\BBC^2$ (or $\mathbb{H}$) which would do the same job.
\end{open}

In the longer perspective, we recall that in the work~\cite{Manton87} back from 1987, the standard approach required calculating the eigenvalues of the Jacobian of the field $U\colon \BBR^3\to SU(2)$ at a fixed time $t\in\BBR$. (The Skyrmion is a static solution.) What are the quantities, corresponding to those eigenvalues, now in terms of the Wronskian matrix?

\section*{Acknowledgements}
The first author is grateful to the Organisers of the 36th International Colloqium on Group Theoretical Methods in Physics (13--17 July 2026 in Valladolid, Spain) for a full week of stimulating discussions and a partial waiver of the registration fee. Special thanks are to profs Popov, Dotsenko, and Toppan. The first author thanks Dr~Lewark (ETHZ) for a discussion of the text. The second author thanks prof.\ T.\ Willwacher for hospitality and fruitful academic exchange. Both authors are extremely grateful to prof.\ Mostovoy for stimulating remarks, constructive criticism, and literature suggestions.


\paragraph{Funding information}
The first author is partially supported by the Bernoulli Institute (via its graduate programme). The stay of the second author at ETH Z\"urich is supported by ETH Z\"urich and the Bernoulli Swiss travel fellowship (CH). 

\begin{appendix}
\numberwithin{equation}{section}

\section[Expansion of the left-current in su(2)]{Expansion of the left-current $L_\mu\in\su(2)$}\label{AppLeftCurrentsu2}

\begin{proof}[Proof of Lemma~\ref{LemLeftCurrentsu2}]
    Write the $SU(2)$\nobreakdash-valued field as $U(t;\bx)=\bigl(\begin{smallmatrix} \alpha & -\bar{\beta} \\ \beta & \bar{\alpha} \end{smallmatrix}\bigr)$, where $\alpha,\beta(t;\bx)\in\BBC$. Then the left-current $L_\mu$ becomes
    \begin{equation}
        L_\mu(t;\bx)=\begin{pmatrix}
            \bar{\alpha}\partial_\mu \alpha + \bar{\beta}\partial_\mu \beta & 
            \bar{\beta}\partial_\mu\alpha - \bar{\alpha}\partial_\mu \beta \\
            \alpha\partial_\mu \beta - \beta \partial_\mu \alpha & 
            \alpha \partial_\mu \bar{\alpha} + \beta \partial_\mu\bar{\beta}
        \end{pmatrix}.
    \end{equation}
    Expansion of $\partial_\mu \bigl(|\alpha|^2+|\beta|^2\bigr)=0$ becomes $a_\mu + \bar{a}_\mu=0$, where $a_\mu=\bar{\alpha}\partial_\mu \alpha + \bar{\beta}\partial_\mu \beta$ and $\bar{a}_\mu$ are the diagonal elements of $L_\mu$. Putting $b_\mu=\alpha\partial_\mu \beta - \beta \partial_\mu \alpha$, it is evident that $-\bar{\beta}$ is the other off-diagonal element of $L_\mu$, hence $L_\mu=\bigl(\begin{smallmatrix}
        a_\mu & -\bar{b}_\mu \\ b_\mu & \bar{a}_\mu
    \end{smallmatrix}\bigr)$, where $a_\mu$ is purely imaginary (as $\bar{a}_\mu=-a_\mu$). A direct calculation shows that $L_\mu^\dagger = -L_\mu$, whence the claim $L_\mu\in\su(2)$ follows.
\end{proof}

Let us express $\cB^\mu$ in terms of the generators $a_\mu\in i\BBR$ and $b_\mu\in\BBC$. 
\begin{prop}
    With $a_\mu$ and $b_\mu$ as in the proof of Lemma~\ref{LemLeftCurrentsu2}, the baryon 4\nobreakdash-current $\cB^\mu$ becomes
    \begin{equation}\label{EqBmuFromTwoGens}
        \cB^\mu = 6\varepsilon^{\mu\!\nu\sigma\!\rho} a_\rho b_\nu \bar{b}_\sigma,
    \end{equation}
    which is a real-valued expression.
\end{prop}
\begin{proof}
    Writing $L_\mu$ in terms of its real coordinates w.r.t.\ the basis $\{i\sigma_a\}$ of $\su(2)$, $L_\mu=i\sigma_a L_\mu^a$, and comparing it with $L_\mu=\bigl(\begin{smallmatrix}
        a_\mu & -\bar{b}_\mu \\ b_\mu & \bar{a}_\mu
    \end{smallmatrix}\bigr)$ yields
    \begin{align}
        L_\mu^1 &= \Imag b_\mu  &  L_\mu^2 &= -\Real b_\mu & L_\mu^3 &= \Imag a_\mu = -ia_\mu.
    \end{align}
    Wherefore the expression for the current $\cB^\mu=\varepsilon^{\mu\!\nu\sigma\!\rho}\tr(L_\nu L_\sigma L_\rho)=12\varepsilon^{\mu\!\nu\sigma\!\rho}L_\nu^1 L_\sigma^2 L_\rho^3$ becomes $\cB^\mu=12i\varepsilon^{\mu\!\nu\sigma\!\rho}\Imag b_\mu \Real b_\sigma a_\rho$ (using $\Imag a_\mu=-i a_\mu$). Now, the anti-symmetrisation relation $b_\nu \bar{b}_\sigma - b_\sigma \bar{b}_\rho = 2i (\Imag b_\nu \Real b_\sigma - \Imag b_\sigma \Real b_\nu)$ lets us write $\cB^\mu = \tfrac{12i}{2i}\varepsilon^{\mu\!\nu\sigma\!\rho} a_\rho b_\nu \bar{b}_\sigma$, \textit{qed}.
\end{proof}

\section{Proof of Lemma~\ref{LemCoordVectorL}}\label{AppProofLemmaL}

\begin{proof}[Proof of Lemma~\ref{LemCoordVectorL}]
    Denote the unit vector of $\vec{\theta}$ by $\hat{\btheta}\ceq(\hat{\theta}^1,\hat{\theta}^2,\hat{\theta}^3)=\vec{\theta}/\theta$. Recall that the re-mapped field is $\bm=(\cos\theta,\sin\theta\, \hat{\bn})=(\cos\theta,\sin\theta[\hat{\theta}^3,-\hat{\theta}^2,\hat{\theta}^1])$. The expressions of the orthogonal vectors from Lemma~\ref{LemOrthogBasis} are
    \begin{alignat*}{7}
        \hat{\be}_1 &= (-\sin\theta\,\hat{\theta}^1, && & & \sin\theta\,\hat{\theta}^2, && & &\sin\theta\,\hat{\theta}^1, && & &\cos\theta), \\
        \hat{\be}_2 &= (-\sin\theta\,\hat{\theta}^2, &&-& &\sin\theta\,\hat{\theta}^1,&&-& &\cos\theta, && & &\sin\theta\,\hat{\theta}^3),\quad \text{and} \\
        \hat{\be}_3 &= (-\sin\theta\,\hat{\theta}^3, && & &\cos\theta, &&-& &\sin\theta\,\hat{\theta}^1,&&-& &\sin\theta\,\hat{\theta}^2).
    \end{alignat*}
    Notice that the dot products $L_i^a=\hat{\be}_a\cdot\partial_i\bm$ all take the form
    \begin{equation}
        L_i^a = \cos\theta\,\partial_i(\sin\theta\, \hat{\theta}^a) - \sin\theta\,\hat{\theta}^a\partial_i(\cos\theta) + \varepsilon_{abc}\sin\theta\,\hat{\theta}^b\partial_i(\sin\theta\, \hat{\theta}^c).
    \end{equation}
    Using $\partial_i(\cos\theta)=-\sin\theta\,\partial_i\theta$ and $\partial_i(\sin\theta\, \hat{\theta}^a)=\cos\theta\, \partial_i\theta\, \hat{\theta}^a + \sin\theta\, \hat{\theta}^a$, we find
    \begin{align}
        L_i^a &= \cos^2\theta\,\hat{\theta}^a\partial_i\theta + \sin\theta\cos\theta\, \partial_i \hat{\theta}^a + \sin^2\theta\,\hat{\theta}^a\partial_i\theta + \varepsilon_{abc} \sin^2\theta\, \hat{\theta}^b \partial_i \hat{\theta}^c \notag{}\\
        &= \hat{\theta}^a\partial_i\theta + \sin\theta\bigl( \cos\theta \partial_i \hat{\theta}^a + \sin\theta (\hat{\btheta}\times \partial_i \hat{\btheta})^a \bigl).
    \end{align}
    Whence we obtain the vector form
    \begin{equation}
        \vec{L}_i = \hat{\btheta} \partial_i \theta + \sin\theta \bigl( \cos\theta\, \partial_i \hat{\btheta} + \sin\theta\, \hat{\btheta} \times \partial_i \hat{\btheta}\bigr).
    \end{equation}
    As $\partial_i \hat{\btheta} \perp \hat{\btheta}$ and $\partial_i \hat{\btheta} \in T_{\hat{\btheta}}\BBS^2$, the vector $\hat{\btheta}\times \partial_i \hat{\btheta} \in T_{\hat{\btheta}}\BBS^2$ is the vector $\partial_i \hat{\btheta}$ rotated by $+90^\circ=\pi/2$ w.r.t.\ a posit.\!-orient.\ orthonormal basis $\hat{\btheta},\hat{\bepsilon}_1,\hat{\bepsilon}_2$ ($\hat{\btheta}\times\hat{\bepsilon}_1=\hat{\bepsilon}_2$) on $\BBR^3\supsetneq T_{\hat{\btheta}}\BBS^2$: $\hat{\btheta}\times \partial_i \hat{\btheta} = R_{\pi/2}\partial_i \hat{\btheta}$. Notice that the sum of the matrices $\cos\theta\,\id_{2\times 2} + \sin\theta\, R_{\pi/2}$ is a rotation by $\theta=|\vec{\theta}|$, denoted $R_\theta$. Therefore, $\vec{L}_i$ becomes $\vec{L}_i=\hat{\btheta}\partial_i\theta + \sin\theta\, R_\theta (\partial_i \hat{\btheta})$ as claimed.

    Let us now express $\vec{L}_i$ as a column vector in the basis $\hat{\btheta},\hat{\bepsilon}_1,\hat{\bepsilon}_2$ on $\BBR^3$: put $\partial_i\hat{\btheta}=p_i\hat{\bepsilon}_1+q_i\hat{\bepsilon}_2$, hence we find $\vec{L}_i=\bigl[\partial_i\theta~ \sin\theta\,R_\theta[p_i~q_i]^\top\bigr]^\top$. Thus $\det\bigl([L_i^a]_{i,a=1,2,3}\bigr)=\det[\vec{L}_1,\vec{L}_2,\vec{L}_3]$ 
    becomes
    \begin{equation}
        \det\bigl([L_i^a]\bigr) = \det \left( \begin{bmatrix}
            1 & 0 & 0\\
            0 & & \\
            0 & \multicolumn{2}{c}{\smash{\raisebox{.5\normalbaselineskip}{$\sin\theta\,R_\theta$}}}
        \end{bmatrix} 
        \begin{bmatrix}
            \partial_1 \theta & \partial_2 \theta & \partial_3 \theta \\
            p_1 & p_2 & p_3 \\
            q_1 & q_2 & q_3
        \end{bmatrix}
        \right) =
        \sin^2\theta\, \det R_\theta ~ \varepsilon^{i\!jk}\partial_i\theta\,p_j\,q_k,
    \end{equation}
    whence the claimed equality~\eqref{EqDetLProvisional} follows (using $\det R_\theta=1$).
\end{proof}


\end{appendix}








\begin{thebibliography}{19}\small 

\bibitem{MultifacSkyrmion}
G. E. Brown and M. Rho, ``\textit{The multifaceted skyrmion}'', World Scientific Publishing Co.\ 
(2010).

\bibitem{Dzhuma2002}
A. S. Dzhumadil'daev, \textit{$n$-Lie structures generated by Wronskians}, Siberian Math.~J. \textbf{46}, 4, 601--612 (2005) \doi{10.1007/s11202-005-0061-7}

\bibitem{Fabbri25}
L. Fabbri, \textit{Euler and Pontryagin currents of the Dirac operator}, J. Phys. A: Math. Theor. \textbf{58}, 025205 (2025) \doi{10.1088/1751-8121/ad9e57}\,, corrigendum (2025) \doi{10.1088/1751-8121/add6ea}

\bibitem{FosterKrusch13}
D. Foster and S. Krusch, \textit{Negative baryon density and the folding structure of the $B=3$ Skyrmion}, J.~Phys.~A: Math. Theor. \textbf{46}, 265401 (2013) \doi{10.1088/1751-8113/46/26/265401}

\bibitem{WronskReview}
L. Gatto and I. Scherbak, \textit{On generalised Wronskians}, EMS Ser. Congr. Rep. Proc. `The Impanga Conf. on Alg. Geo.', 
Eur.\ Math.\ Soc., Z\"{u}rich (2012) \doi{10.4171/114-1/9}

\bibitem{HanlonWachs}
P. Hanlon, M. Wachs, \textit{On Lie $k$-algebras}, Adv. Math. \textbf{113}, 2, 206--236 (1995)
\doi{10.1006/aima.1995.1038}

\bibitem{MantonRatmapSkyrmions}
C. J. Houghton, N. S. Manton, P. M. Sutcliffe, \textit{Rational maps, monopoles and Skyrmions}, Nucl. Phys. B. \textbf{510}, 3, 507--537 (1997) \doi{10.1016/S0550-3213(97)00619-6} 

\bibitem{FinDim}
M. G. \c{K}\={e}ni\c{n}\v{s} and A. V. Kiselev, \textit{Explicit class of finite-dimensional polynomial algebras with Wronskians over $\mathbb{R}^d$ as $N$\nobreakdash-ary Lie brackets: beyond $\mathfrak{sl}(2)$}, (arXiv preprint) (2026)
\doi{10.48550/arXiv.2605.27305}

\bibitem{FiboWron}
M. G. \c{K}\={e}ni\c{n}\v{s} and A. V. Kiselev, \textit{Iterate Wronskians over $\mathbb{R}^d$ as $N$\nobreakdash-ary brackets on $\mathbb{R}[x^1,\cdots,x^d]$: the $N$\nobreakdash-bonacci numbers bound the highest total degrees}, (arXiv preprint) (2026)
\doi{10.48550/arXiv.2607.26039}

\bibitem{ForKac}
A. V. Kiselev, \textit{On associative Schlessinger–Stasheff algebras and Wronskian determinants}, J.~Math. Sci. \textbf{141}, 1, 1016–1030 (2007), \doi{10.1007/s10958-007-0028-2} 

\bibitem{PRG25Wronsk}
A. V. Kiselev, \textit{Wronskians as $N$-ary brackets in finite-dimensional analogues of~$\mathfrak{sl}(2)$}, J.~Phys.: Conf. Ser. \textbf{3152}, 
012044 (2025) \doi{10.1088/1742-6596/3152/1/012044}

\bibitem{YER25}
A. V. Kiselev, \textit{Jacobi identities for Wronskian determinants
over multidimension}, Bulgarian J.~Phys. \vol{52}-s1, 
102--107 (2025) \doi{10.55318/bgjp.2025.52.s1.102}

\bibitem{LeVeque1956}
W. J. LeVeque, ``\textit{Topics in number theory}'', Vol.~\textbf{2},
Addison--Wesley Publ. Co. (1956). 

\bibitem{30Anniv}
V. G. Makhan'kov, Yu. P. Rybakov, V. I. Sanyuk, \textit{The Skyrme model and strong interactions (On the 30th anniversary of the creation of the Skyrme model)}, Sov. Phys. Usp. \textbf{35}, 2, 55–84 (1992)
\doi{10.1070/PU1992v035n02ABEH002214}

\bibitem{Manton87}
N. S. Manton, \textit{Geometry of Skyrmions}, Commun. Math. Phys. \textbf{111}, 469--478 (1987), \doi{10.1007/BF01238909}

\bibitem{Manton}
N. S. Manton, \textit{Classical Skyrmions -- static solutions and dynamics}, Math. Meth. Appl. Sci. \textbf{35}, 1188--1204 (2012) \doi{10.1002/mma.2512}

\bibitem{MilasQSeries}
A. Milas, E. Mortenson, and K. Ono, \textit{Number theoretic properties of Wronskians of Andrews-Gordon series}, Int. J. Number Theory \textbf{4}, 2, 323--337 (2008) \doi{10.1142/S1793042108001377}

\bibitem{MonastyrskyTGF}
M. Monastyrsky, ``\textit{Topology of gauge fields and condensed matter}'', Springer New York (1993) \doi{10.1007/978-1-4899-2403-2}

\bibitem{PolyaGFS}
A. M. Polyakov, ``\textit{Gauge fields and strings}'', Contemporary Concepts in Physics \textbf{3}, Harwood Academic Publishers GmbH 
(1987).






\bibitem{Sutcliffe23}
P. Sutcliffe, \textit{A Skyrme model with novel chiral symmetry breaking}, SIGMA \textbf{19}, 051 (2023), \doi{10.3842/SIGMA.2023.051}



\bibitem{Schmidt1980}
W. M. Schmidt, ``\textit{Diophantine approximation}'', Lect. Notes in Math. \textbf{785}, Springer, Berlin (1980), \doi{10.1007/978-3-540-38645-2}

\bibitem{Wolsson1989a}
K. Wolsson, \textit{Linear dependence of a function set of $m$ variables with vanishing generalized Wronskians}, Linear Algebra Appl., \textbf{117}, 73--80 (1989) \doi{10.1016/0024-3795(89)90548-X}

\bibitem{RajaramanSI}
R. Rajaraman, ``\textit{Solitons and instantons: an introduction to solitons and instantons in quantum field theory}'', North--Holland Publishing Co. 
(1982).

\bibitem{SchwarzTopolPhys}
A. S. Schwarz, ``\textit{Topology for physicists}'', Springer Berlin, Heidelberg (1994) \doi{10.1007/978-3-662-02998-5}

\bibitem{SkyrmeSkyrmions}
T. H. R. Skyrme, 
\textit{A non-linear theory of strong interactions}, Proc. R. Soc. A \textbf{247}, 1249, 260–278 (1958), \doi{10.1098/rspa.1958.0183}

T. H. R. Skyrme, 
\textit{A unified model of $K$- and $pi$-mesons}, Proc. R. Soc. A \textbf{252}, 1269, 236–245 (1959) \doi{10.1098/rspa.1959.0149}

T. H. R. Skyrme, 
\textit{A non-linear field theory}, Proc. R. Soc. A \textbf{260}, 1300, 127--138 (1961) \doi{10.1098/rspa.1961.0018}

T. H. R. Skyrme, 
\textit{Particle states of a quantised meson field}, Proc. R. Soc. A \textbf{262}, 1309, 237--245 (1961) \doi{10.1098/rspa.1961.0115}

\bibitem{Stasheff2019Review}
J. Stasheff, \textit{$L_\infty$ and $A_\infty$ structures}, High. Struct. \textbf{3}, 1, 292--326 (2019) \doi{10.21136/HS.2019.07}


\end{thebibliography}


\end{document}